%% file: ARXIV_AA_29629_am.tex
\documentclass[structabstract]{aa}  
\usepackage{graphicx}
\usepackage{graphicx,color,layout}
\usepackage{epstopdf}
\usepackage{txfonts}
\usepackage{natbib}
\usepackage{lscape}

\begin{document}

   \title{The AMBRE project: Iron-peak elements in the solar neighbourhood\thanks{Based on observations collected at ESO telescopes under the 
AMBRE programme.
}\fnmsep \thanks {Table \ref{tab:all_stars} is only available in electronic form
at the CDS via anonymous ftp to cdsarc.u-strasbg.fr (130.79.128.5)
or via http://cdsweb.u-strasbg.fr/cgi-bin/qcat?J/A+A/}
}

   \author{\v{S}.~Mikolaitis
          \inst{1,2}
          \and
          P.~de~Laverny
          \inst{2}
          \and  
          A.~Recio--Blanco
          \inst{2}
          \and
          V.~Hill
          \inst{2}
          \and
          C.C.~Worley 
          \inst{3}
          \and
          M.~de~Pascale 
          \inst{4}
          }

   \institute{
   Institute of Theoretical Physics and Astronomy, Vilnius University, Saul\.{e}tekio al. 3, LT-10257, Vilnius, Lithuania\\
              \email{Sarunas.Mikolaitis@tfai.vu.lt} \\
                       \and
         Universit\'e C\^ote d'Azur, Observatoire de la C\^ote d'Azur, CNRS, Laboratoire Lagrange, Bd de l'Observatoire, CS\_34229, 06304 Nice cedex 4, France\\
                              \and
        Institute of Astronomy, University of Cambridge, Madingley Road, Cambridge CB3 0HA, United Kingdom \\
        \and
        INAF - Padova Observatory, Vicolo dell'Osservatorio 5, 35122 Padova, Italy
             }

   \date{Received September 01, 2016; accepted November 24, 2016; published March 22 2017}

\authorrunning{Name Surname} 

  \abstract
{
  The pattern of chemical abundance ratios in stellar populations of the Milky Way is a fingerprint of the Galactic chemical history. In order to interpret such chemical fossils of Galactic archaeology, chemical evolution models have to be developed. However, despite the complex physics included in the most recent models, significant discrepancies between models and observations are widely encountered.
   }
{
  The aim of this paper is to characterise the abundance patterns of five iron-peak elements (Mn, Fe, Ni, Cu, and Zn) for which the stellar origin and chemical evolution are still debated.
  }
{
  We automatically derived iron peak (Mn, Fe, Ni, Cu, and Zn) and $\alpha$ element (Mg) chemical abundances for 4\,666 stars, adopting classical LTE spectral synthesis and 1D atmospheric models. Our observational data collection is composed of high-resolution, high signal-to-noise ratios HARPS and FEROS spectra, which were previously parametrised by the AMBRE project.
   }
{
  We used the bimodal distribution of the magnesium-to-iron abundance ratios to chemically classify our sample stars into different Galactic substructures: thin disc, metal-poor and high-$\alpha$ metal rich, high-$\alpha,$ and low-$\alpha$ metal-poor populations.
  Both high-$\alpha$ and low-$\alpha$ metal-poor populations are fully distinct in Mg, Cu, and Zn, but these substructures are statistically indistinguishable in Mn and Ni.
  Thin disc trends of [Ni/Fe] and [Cu/Fe] are very similar and show a small increase at supersolar metallicities. Also, both thin and thick disc trends of Ni and Cu are very similar and indistinguishable. Yet, Mn looks very different from Ni and Cu. [Mn/Fe] trends of thin and thick discs actually have noticeable differences: the thin disc is slightly Mn richer than the thick disc. The [Zn/Fe] trends look very similar to those of [$\alpha$/Fe] trends. The typical dispersion of results in both discs is low ($\approx$0.05 dex for [Mg, Mn, and Cu/Fe]) and is even much lower for [Ni/Fe] ($\approx$0.035~dex).
}
   {
  It is clearly demonstrated that Zn is an $\alpha$-like element and could be used to separate thin and thick disc stars. Moreover, we show that the [Mn/Mg] ratio could also be a very good tool for tagging Galactic substructures. From the comparison with Galactic chemical evolutionary models, we conclude that some recent models can partially reproduce the observed Mg, Zn, and, Cu behaviours in thin and thick discs and metal-poor sequences. Models mostly fail to reproduce Mn and Ni in all metallicity domains, however, models adopting yields normalised from solar chemical properties reproduce Mn and Ni better, suggesting that there is still a lack of realistic theoretical yields of some iron-peak elements. The very low scatter ($\approx$0.05 dex) in thin and thick disc sequences could provide an observational constrain for Galactic evolutionary models that study the efficiency of stellar radial migration.
   }


\maketitle

\section{Introduction}

The pattern of chemical abundances directly reflect the history of a stellar population. The main contributors to the interstellar medium chemical evolution are believed to be asymptotic giant branch (AGB) stars and type-Ia and type-II supernovae. 
The final picture of stellar abundances in the Galaxy is thus predetermined by a complex interplay of several mechanisms: a supernovae type that was predominant in the region of the stellar nursery, the typical mass of type-II supernovae when stars formed, the timing of the mechanism that switched on the type-Ia supernovae, and the amount of contamination in the region by supernovae and AGB ejecta. This information is important to unlocking the knowledge of the formation and evolution histories of our Galaxy. Among the numerous observational properties that are usually collected to
describe the Galactic populations, one of the most powerful properties consists in collecting chemical abundance ratios between different key 
elements. Then, such ratios in the different Galactic components have to be interpreted together with other signatures 
(such as kinematical signatures, for instance) by chemical models of the Milky Way.
Obviously, the solar vicinity is the most studied region of the Milky Way and several chemical evolution models have been developed to explain 
the chemical composition of the local disc (e.g. \citealt{Chiappini2003a,Chiappini2003b,Francoise2004}). However, most
of these studies typically modelled the solar vicinity as one merged system without reproducing the chemical patterns of Galactic substructures. Nevertheless, there are several more recent models \citep{Kobayashi2006, Kobayashi2011, Romano2010, Kubryk2015} that aimed to reproduce most of the chemical abundance patterns of the four main Galactic components (bulge, halo, thin, and thick discs) for elements up to Z=30.
Although such Galactic chemical evolution models are very complex and advanced, one could still see that part of the chemical patterns 
found in the solar neighbourhood are still not well understood.

With this paper, we aim to better describe the behaviour of the manganese, nickel, copper, and zinc elements in the Galaxy. We chose these iron-peak chemical species  since they are key products of stellar nucleosynthesis.

Manganese ($^{55}$Mn) is produced during thermonuclear explosive silicon burning and $\alpha$-rich freeze-out via synthesis of $^{55}$Co. These
reactions are active in type-Ia (\citealt{Bravo2013}) and type-II (\citealt{Woosley1995}) supernovae, although these reactions are thought to be produced more by type-Ia supernovae than by hypernovae or type-II supernovae (\citealt{Iwamoto1999, Kobayashi2009}). However, it is still under debate whether the manganese-to-iron trends behave differently in thin and thick discs at similar metallicities, 
probably revealing different evolution histories (\citealt{Feltzing2007, Hawkins2015, Battistini2015}).
Nickel is also produced by the silicon burning process. However, yields from type-Ia and type-II supernovae are comparable. Thus nickel behaves almost like iron from an observational point of view. However,  it is still very difficult for Galactic models to match the observed nickel behaviour in the Milky Way.
Copper is an odd iron-peak element and its origin is still not well understood. For example, it is thought that solar copper is mostly the outcome of weak $s$ and $r$ processes (\citealt{Bisterzo2004}). However, there are other nucleosynthesis channels known for copper, but their relative contributions are under debate (\citealt{Bisterzo2006}).  
Finally, in massive stars most of the zinc should have been produced by similar nucleosynthesis channels such as copper, but with different contribution ratios. $^{64}Zn$ is mostly produced from $\alpha$-rich freeze-out in $\gamma$ winds and ($^{66,67,68,70}Zn$) is produced by the \textit{sr} process. Other contributions are marginal (\citealt{Bisterzo2004, Bisterzo2006}). The zinc-to-iron ratio pattern should resemble that of the $\alpha$ elements. However, such a behaviour is still not explicitly modelled or confirmed by observations based on large Galactic samples.

Indeed, large and homogeneous samples of Mn, Ni, Cu, and Zn abundances are required in order to better understand the evolution of these elements in our Galaxy and to test the quality of chemical evolution models. Although there are already some significant samples of Mn, Ni, Cu, and Zn abundances derived from high-resolution spectra, the size and completeness of these samples vary a lot from element to element. 
For instance, the largest uniform samples of manganese and nickel abundances were provided by \citeauthor{Adibekyan2012} (\citeyear{Adibekyan2012}; 1\,111 dwarf stars), \citeauthor{Bensby2014} (\citeyear{Bensby2014}; 714 dwarfs), \citeauthor{Battistini2015} (\citeyear{Battistini2015}; 596 dwarfs), and \citeauthor{Hawkins2015} (\citeyear{Hawkins2015}; 3\,200 giants). However, copper and zinc have been given less attention in observational studies. The largest samples of Cu were studied by \citeauthor{Reddy2008} (\citeyear{Reddy2008}; uniform sample of 181 stars) and \citeauthor{Bisterzo2006} (\citeyear{Bisterzo2006}; non-uniform collection of abundances for more than 300 stars).
There exists only one uniform and large sample reporting zinc abundances \citep[][714 dwarf stars]{Bensby2014}, whereas \citet{Bisterzo2004} collected a non-uniform sample of more than 300 stars. 

In order to increase the size of observed samples for iron-peak elements by about one order of magnitude, in this article we present  a very large catalogue of homogeneous magnesium, manganese, iron, nickel, copper, and zinc elemental abundances derived for 4,666 slow-rotating stars.
Our sample contains stars in various evolutionary stages and covers a large metallicity domain in the different Galactic components. This study is part of the Galactic Archaelogy AMBRE project relying on archived ESO spectra (\citealt{Laverny2013}). In Sect.~\ref{sec:analysis}, we describe the observational data sample and the method adopted to derive the chemical abundances. Sect.~\ref{sec:definition_of_components} is devoted to tagging the analysed stars to the different Galactic components based on their
enrichment in magnesium. In Sect.~\ref{sec:IroninSolarN}, we describe the behaviour of the iron-peak elements in these different Galactic disc components. In Sect.~\ref{sec:models_comp}, we compare our observational results with recent Galactic chemical evolution models. Finally, we summarise and conclude our work in Sect.~\ref{sec:conclusion}.

\section{The AMBRE catalogue of iron-peak elements}
\label{sec:analysis}
\subsection{Stellar sample and atmospheric parameters}
For the present study, we adopted the  AMBRE project data that are described in \citet{Laverny2013} and consist in the automatic parametrisation of large sets of ESO high-resolution archived spectra. 
Within the AMBRE project,the stellar atmospheric parameters 
(in particular, $T_{\rm eff}$, ${\rm log}~g$, [M/H], and [$\alpha$/Fe] together with their associated errors)
were derived using the MATISSE algorithm (\citealt{Blanco2006}).
For the present study, we adopted the data of the FEROS and HARPS spectrographs (hereafter AMBRE:FEROS and AMBRE:HARPS samples).
The complete parametrisation of the AMBRE:FEROS spectra has been 
performed by \citet{Worley2012} (hereafter W12) for 6,508 spectra, whereas
the 90,174 AMBRE:HARPS spectra have been fully parametrised by \citet{dePascale2014} (hereafter dP14).

For some stars, these two samples (and particularly the HARPS sample) contain a large number of repeated observations.
The actual number of targeted stars is thus smaller than the number of parametrised spectra.
As already shown in W12, the number of AMBRE:FEROS stars is actually 3\,087 as defined by their coordinates. 
For the AMBRE:HARPS sample, the actual number of stars 
has been calculated by, first, analysing the spectra in a 3D space (RA, DEC, and radial velocity).
For that purpose, we selected every spectra found within a radius of 5" and look for clusters
in the radial velocity (Vrad). We defined independent clusters as spectra with Vrad departing by 
more than 2$\sigma$ from the mean Vrad value of the cluster.
Then, we checked the atmospheric parameters to disentangle stars within a cluster.
We thus made the so-called single linkage cluster analysis for well-separated clusters to search for the statistical clusterisation of solutions in a 4D space ($T_{\rm eff}$, ${\rm log}~g$, [M/H], and [$\alpha$/Fe]). 
If all 4D parameters of one cluster of solutions were different from another cluster by at least 2$\sigma$ in the parameter values, we accepted this solution as a single star.
In such a way, we found 5,049 individual stars in the AMBRE:HARPS sample. 
Moreover, if a given star did not have a single observation with CHI2\_FLAG=0 (quality flag of the fit between the observed and reconstructed synthetic spectra with the AMBRE parameters, zero meaning \textit{very good fit}), this star was omitted from further analysis.
This procedure led to a AMBRE:HARPS sample with 4,355 stars and 
to a AMBRE:FEROS sample with 1\,925 stars.

For the chemical analysis presented in Sect.~\ref{sec:abundanceanalysis} and in case of repeats
within one of these two samples, we selected one spectrum per star by favouring the best combination of AMBRE CHI2 and S/N values. 
The CHI2 is the log($\chi^2$) fit between the observed spectrum and the synthetic spectrum reconstructed with the AMBRE atmospheric parameters. The S/N is a signal-to-noise ratio estimated by the AMBRE pipeline (see Table~A.1 in W12 and Table~2 in dP14 for details). 
In case of multiple observations of the same star (ten or more), we first selected five spectra with the best S/N. Then, among these spectra, we selected the spectrum with the highest CHI2 value. If the number of repeats is rather small ($\sim$10), we simply selected the spectrum with the best S/N.

 \begin{figure*}[htb]
   \advance\leftskip 0cm
   \includegraphics[scale=0.15]{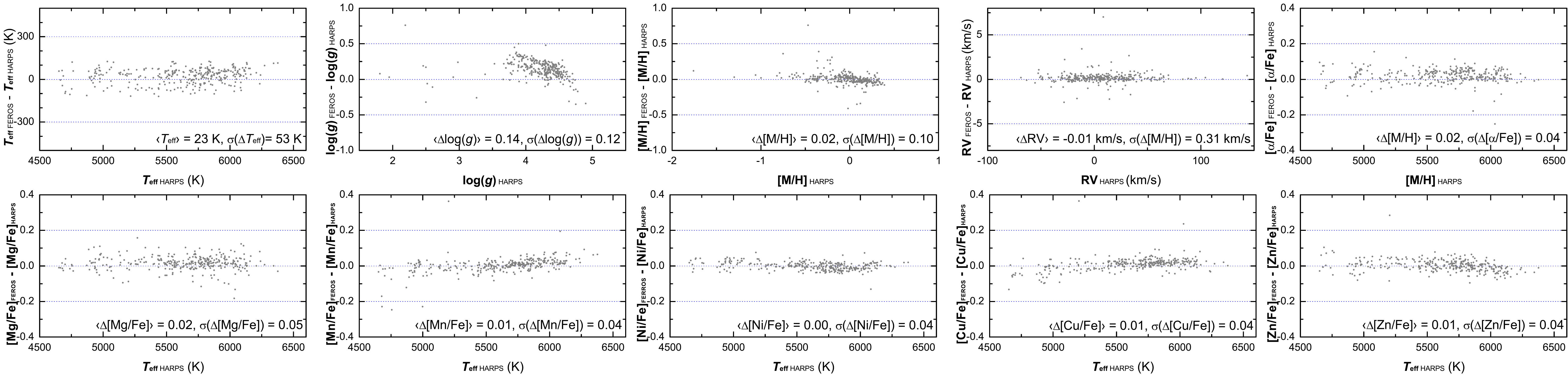}
  \caption{
Comparison between the stellar parameters derived by the AMBRE:FEROS and the AMBRE:HARPS pipeline (top five plots); comparison between [El/Fe] abundance ratios (lower five plots) for 320 stars in common between the final AMBRE:FEROS and AMBRE:HARPS samples.
  }
  \label{fig:FEROSvsHARPS}
  \end{figure*}

Finally, we also found that there are 320 stars in common between the final AMBRE:FEROS and AMBRE:HARPS samples; this was checked by coordinates, radial velocities, and atmospheric parameters. The comparison between FEROS and HARPS stellar parameters and abundances for 320 stars in common is shown in Fig.~\ref{fig:FEROSvsHARPS}. 
Therefore, even if a small dispersion and bias are present for some stellar parameters, they do not affect the derived [El/Fe] abundance ratios.
We then considered only the HARPS spectra for our final analysis since they usually have higher S/N values and higher spectral resolution.
We therefore have a final catalogue of 5\,960 (4\,355 HARPS and 1\,605 FEROS) individual slow-rotating stars (and spectra) homogeneously parametrised by the AMBRE project.

\subsection{Chemical abundance analysis}
\label{sec:abundanceanalysis}
\subsubsection{Adopted methodology}
The method used to derive individual chemical abundances of Mg, Mn, Fe, Ni, Cu, and Zn
is described in \citet{Mikolaitis2014}. Here we briefly describe the most relevant information. We adopted the
AMBRE atmospheric parameters ($T_{\rm eff}$, ${\rm log}~g$, $\langle{\rm{[M/H]}}\rangle$, and [$\alpha$/Fe]) to derive the chemical abundances.
We use the notation [M/H] for the metallicity of the atmospheric model and the value derived by the AMBRE Project. 
For the neutral and ionised iron abundances, we use the notations [\ion{Fe}{I}/H] and [\ion{Fe}{II}/H]. Similarly, 
[$\alpha$/Fe] refers to the $\alpha$ to iron ratio of the atmospheric model derived by AMBRE.
We computed the microturbulent velocity using the fixed function of $T_{\rm eff}$, ${\rm log}~g$, and [M/H] as defined within
the Gaia-ESO Survey (GES) (Bergemann et al., in preparation). We also kept the original spectral resolution for the analysis,
i.e. around 48,000 and 120,000 for FEROS and HARPS, respectively. These spectral resolutions
are much larger than that
adopted by the AMBRE Project for the stellar parametrisation.

\begin{table}
\caption{Lines ($\lambda(\AA)$) used for the analysis.
}
\resizebox{\columnwidth}{!}{%
\begin{tabular}{llllllllllllllllllllll}
\hline\hline
\ion{Mg}{I} \\
\hline

5167.3 & 5172.6 & 5183.6 & 5528.4 & 5711.0 & 6318.7 \\ 
6319.2 & 6319.4 & 
\\  \hline
\ion{Mn}{I} \\
4783.4$^B$ & 4823.5$^B$ & 5004.9$^{BW}$ & 5117.9$^J$ & 5255.3$^J$ & 5394.7$^D$ \\ 
5407.4$^{De}$ & 5420.4$^{De}$ & 5432.5$^D$ & 5516.8$^{De}$ & 6013.5$^{H}$ & 6016.7$^{H}$ \\ 
6021.8$^{H}$ & 6440.9$^{H}$ &        \\
\hline
\ion{Ni}{I} \\
4811.9  &       4814.5  &       4873.4  &       4953.2  &       4976.1  &       4976.3  \\
4976.6  &       5003.7  &       5035.3  &       5137.0  &       5157.9  &       5424.6  \\
5435.8  &       5578.7  &       5587.8  &       5709.5  &       5846.9  &       6108.1  \\
6128.9  &       6176.8  &       6204.6  &       6223.9  &       6230.0  &       6327.5  \\
6378.2  &       6384.6  &       6414.5  &       6482.7  &       6643.6  &       6767.7  \\
\hline
\ion{Cu}{I} \\
5105.5$^{F}$ & 5153.2 & 5218.2$^{He}$ & 5220.1$^{He}$ & 5700.2$^{Be}$ & 5782.1$^{Be}$ \\ 
\hline
\ion{Zn}{I} \\
4722.1 & 4810.5 & 6362.3  \\
\hline
 \label{tab:TABLE_LINES}
\end{tabular}%
}
\\
HFS data from: $^B$\citet{Brodzinski1987}, $^{BW}$\citet{Blackwell-Whitehead2005a}, $^J$\citet{Johann1981}, $^D$\citet{Davis1971}, $^{De}$\citet{Dembczynski1979}, $^{H}$\citet{Handrich1969}, $^{F}$\citet{Fischer1967}, $^{He}$\citet{Hermann1993}, $^{Be}$\citet{Bergstrom1989}
\\
\end{table}  

As for the atomic line selection of the considered chemical species, 
we first look for lines that can be found in both spectral domains covered by the FEROS
and HARPS instruments. Then, to consider the best possible line data, we looked at the
lines provided by the Gaia-ESO Survey line list group (\citealt{Heiter2015}). Atomic lines\ that were selected for the abundance analysis are presented in Table~\ref{tab:TABLE_LINES}.  
Iron lines are too numerous (139 in total) to be presented in Table~\ref{tab:TABLE_LINES}, but we confirm that these lines are typically used in stellar spectroscopy since they are well observed in solar or Arcturus spectra (see \citealt{Sousa2014} for instance).
Some types of stars were slightly sensitive to molecular absorption, thus it was necessary to use molecular line lists kindly provided by T.~Masseron for $\rm{C}_{2}$~\citep[][]{Brooke2013,Ram2014}, CN~\citep[][]{Sneden2014}, OH~(Masseron, Priv. comm.), MgH~(Masseron, Priv. comm.), CH~\citep[][]{Masseron2014}, CaH~(Plez, Priv. comm.), NH~(Masseron, Priv. comm.), SiH~\citep[][]{Kurucz1993}, and FeH~\citep[][]{Dulick2003}.

Regarding the normalisation of the observed spectra, our chemical analysis pipeline determines the continuum level in two steps. First, we employed the DAOSPEC tool (\citealt{Stetson2008}) to estimate the continuum function over the whole spectrum. We used this function as a first guess of the continuum. Then we adjusted the continuum locally in the region ($\pm 5\AA$) around every line of interest. This was carried out by selecting the possible line-free zones of the normalised synthetic spectrum, defined as regions where the intensity of the synthetic spectrum is depressed by less than 0.02\%. If the possible line-free zones are too narrow or do not exist for certain types of stars, we iteratively searched for the possible less contaminated zones in the synthetic spectrum. 
We adopted interpolated MARCS (\citealt{Gustafsson2008}) model atmospheres and the version v12.1.1 of the spectrum synthesis code TURBOSPECTRUM (\citealt{Alvarez1998}) for all spectral synthesis.

Then, the abundance analysis was divided into two independent steps after having performed the radial velocity correction, adopting the AMBRE Vrad. 

First, we estimated the average line broadening for each
star; the spectral line broadening is mainly caused by stellar
rotation ($v{\rm sin} i$). However, lines are also broadened by velocity fields
in the stellar photosphere known as macroturbulence ($v_{\rm mac}$). Unfortunately, disentangling the rotational profile from the radial-tangential $v_{\rm mac}$ profile is difficult, leading to a degeneracy between both quantities measured (\citealt{Doyle2014}). Thus, we assume in the following that the line broadening parameter corresponds to the rotational velocity keeping in mind that there is a small influence of the $v_{\rm mac}$ profile. 
We note that the AMBRE Project focuses on slow-rotating stars only (up to $\sim$20~${\rm km~s}^{-1}$).
The $v{\rm sin} i$ of each star was measured together with their iron [FeI/H] and [FeII/H] abundances, which could differ slightly from the AMBRE mean metallicity [M/H]. We then computed the individual chemical abundances adopting the derived $v{\rm sin}i$. 

The $v{\rm sin}i$ was adjusted iteratively by minimising the $\chi^2$ of the fit between the synthetic spectrum and observed spectrum
around some specific lines.
We chose to adjust $v{\rm sin}i$ using neutral iron lines (up to 139 are available in the analysed spectral domains) because (i) we were able to adopt the AMBRE metallicity as a first guess of the iron abundance, and (ii) iron lines are the most numerous. 
First, for a given spectrum and iron line, we created a grid of synthetic spectra with different $v{\rm sin}i$ values
(from $v{\rm sin}i=0.0$ to 20.0 ${\rm km~s}^{-1}$ with a step 0.1 ${\rm km~s}^{-1}$) and searched for the best fit (lowest $\chi^2$). 
Then, we adopted this chosen $v{\rm sin}i$ as a first guess to compute the iron abundance of the specific line; see below for a short description of the methodology adopted for the chemical analysis. 
We repeated those two steps until convergence on both $v{\rm sin}i$ and [Fe/H] was reached for the specific iron line studied. 
Finally, we computed the mean and scatter of $v{\rm sin}i$, excluding iron lines that were too weak as defined by their line depth (i.e. 
excluding lines with a line depth smaller than twice the noise level). We point out that the number of lines available for this $v{\rm sin}i$ 
estimation varies from star to star. On average, we used 98 iron lines; this number varies from 29 in very metal-poor stars to 139
in metal-rich, high S/N spectra. The derived iron abundances ([FeI/H] and [FeII/H]) are presented in Sect. \ref{sec:catalog} together
with their comparison with the AMBRE [M/H] mean metallicity.

 \begin{figure}[htb]
   \advance\leftskip 0cm
   \includegraphics[scale=0.25]{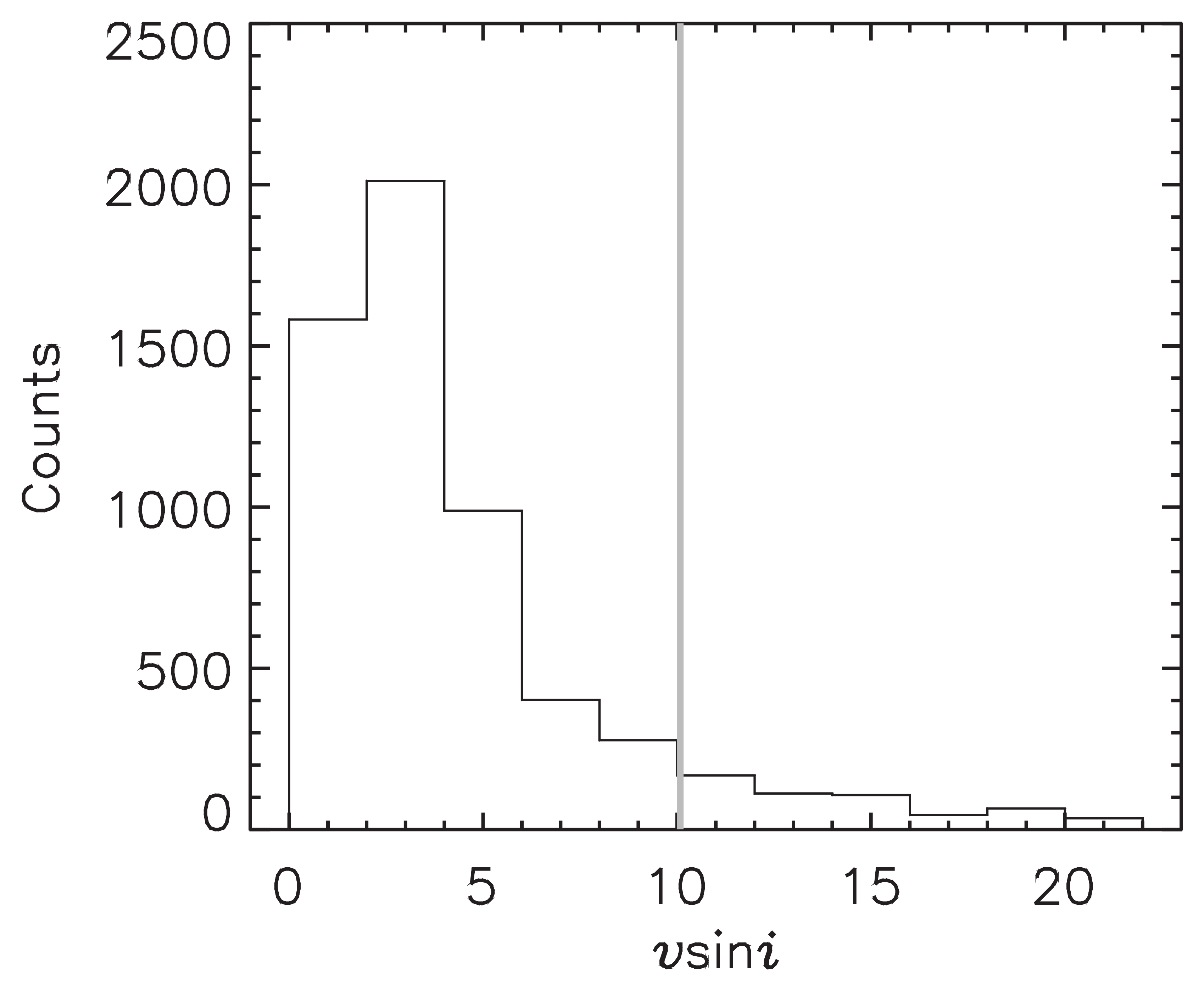}
  \caption{
Rotational velocity distribution in bins of $\Delta v{\rm sin} i$ = 2~km/s. The grey vertical line indicates the selection cut-off at $v{\rm sin} i$ = 10~km/s. 
  }
  \label{fig:rotation_histograms}
  \end{figure}
  
As shown in Fig.~\ref{fig:rotation_histograms}, most of the sample stars are slow rotators ($v{\rm sin}i < 10{\rm km~s}^{-1}$ for 86\% of the sample, $\langle v{\rm sin}i \rangle = 5.1{\rm km~s}^{-1}$). This confirms the assumption adopted within the
AMBRE Project to reject spectra with rotational velocity that is too high (actually, high values of the full width at half maximum (FWHM) of the cross-correlation function (CCF) between the spectra and binary masks), with a limit around $v{\rm sin}i \sim 15~{\rm km~s}^{-1}$.

In the following, and in order to consider only spectra with accurate atmospheric parameters 
derived by the AMBRE pipeline that relies on a grid of non-rotating spectra, we performed
the chemical analysis only for the stars with $v{\rm sin}i < 10~{\rm km~s}^{-1}$.
This very strict criterion rejected 987 stars from the original sample leading
to a final sample of 4\,973 stars (with FEROS or HARPS spectra).

For these stars, we determined the LTE \ion{Mg}{I}, \ion{Mn}{I}, \ion{Ni}{I}, \ion{Cu}{I}, and \ion{Zn}{I} abundances (in addition to \ion{Fe}{I} and \ion{Fe}{II}) by the same line-fit method as in \citet{Mikolaitis2014} using previously derived $v{\rm sin}i$ and the AMBRE atmospheric parameters ($T_{\rm eff}, {\rm log}~g$, ${\rm [M/H]}$, and ${\rm [\alpha/Fe]}$).
We point out that because of the lack of NLTE correction tables available for all the studied species, all forthcoming discussions are based on the derived LTE abundances.
However, some possible NLTE deviations have already been estimated in previous works. For instance, \citet{Mishenina2015} observed average LTE-NLTE abundance variations of 0.01$\pm$0.04 and 0.02$\pm$0.04~dex for the thin and thick discs, respectively. Our LTE determinations for the Mn abundance can be accepted within the given uncertainty of 0.1 dex and the NLTE effect for this element cannot affect our discussion about thin and discs. We reach a similar conclusion about the NLTE effect for Cu abundances since \citet{Shi2014} and \citet{Yan2016} have shown that the NLTE-LTE deviations for typical thin and thick disc stars are generally small (up to 0.1~dex for [Fe/H]$\approx$-1.0~dex). Moreover, \citet{Yan2016} have shown in their Fig.~2 and similar samples as ours that the Cu NLTE effects should not affect our discussion.
Also, we point out, that the \ion{Zn}{I} 6,362 line suffers from \ion{Ca}{I} auto-ionisation broad absorption feature as identified by \citet{Mitchell1965}. However, \citet{Chen2004} showed that
differentially with respect to the Sun, the largest effect from this auto-ionisation line would then be a decrease of [Zn/Fe] for the most metal-poor stars by only about 0.02~dex. Typically, our combined errors are significantly larger than this possible error contribution, which has thus been neglected in the present analysis.

\subsubsection{Error estimation for the chemical abundances}
\label{sec:errors}
The different sources of error impacting the chemical abundances are described in \citet{Mikolaitis2014}.  We briefly discuss these
in the following. 

The first category of errors includes those affecting a single line (e.g. random errors of the line fitting or continuum placement). 
The second category includes the errors affecting all lines. One way (see also the item 1 below) to evaluate this sort of error is to select a statistically significant set of spectra of the same object ({\it repeats})
with different S/N, and then to evaluate the dispersion of the results. 
Another way is to follow the line-to-line scatter (see also the item 2 below).
Finally (see also the item 3), we studied the propagation of the errors from the atmospheric parameters ($T_{\rm eff}, {\rm log}~g$, ${\rm [M/H]}$, and $v_{\rm t}$) to the chemical abundances. These error analyses are presented below.

\input{tab-montecarlo.tex}

\begin{enumerate}
\item
  \textit{Error evaluation using object repeats.} The present project benefits from the large number of repeated observations of the same star (particularly for the HARPS sample). As a consequence, we are able to select a large number of spectra targeting 
the same stars observed during different observational conditions and covering a large domain of S/N values. 
The selected stars (dwarf and giants, metal-rich and poor), their number of available spectra and the estimated 
dispersions ($\sigma$) on the derived chemical abundances are presented in Table~\ref{tab:Montecarlo}.
We point out that the effects induced by differences in the continuum normalisation are included in the estimated
dispersions. One can see that the estimated dispersions are always very small (a couple of hundredths of dex) for every tested
star and any chemical species.
\item
\textit{Error evaluation according line-to-line scatter.} We provide such scatter estimates for all elements in Table~\ref{tab:Scatter}, 
where $\langle\sigma\rangle$ is the mean and $\sigma_{\rm{max}}$ is the maximum of the standard deviation for a given element. These errors are very small and we adopted the values of Table~\ref{tab:Scatter} ($\langle \sigma \rangle$) in the estimation of the total error budget.
\item
\textit{Propagation of the uncertainties on the main atmospheric parameters.}
The typical errors on the atmospheric parameters are provided by the AMBRE Project. For the AMBRE:HARPS sample (\citealt{dePascale2014}) they are:
${ \Delta T_{\rm eff} }\approx$93~K, ${ \Delta \log g }\approx$0.26~dex, and $\Delta {\rm [M/H]}\approx$0.08~dex.
For the AMBRE:FEROS sample (\citealt{Worley2012}) they are:
${ \Delta T_{\rm eff} }\approx$120~K, ${ \Delta \log g }\approx$0.37~dex (if ${\log g}<$3.2 ), ${ \Delta \log g }\approx$0.20~dex (if ${\log g}\geq$3.2), and $\Delta {\rm [M/H]}\approx$0.10~dex. 
The adopted errors of ${ \Delta v_{\rm t} }$ are $\approx$0.3~m$/$s for both samples.
The impact of errors of the atmospheric parameters on the derived abundances are reported in Table~\ref{tab:Sensitivity}. We first provide the typical errors (columns 2~to~5) from errors on $T_{\rm eff}$, ${\rm log}~g$, ${\rm [M/H]}$, and $v_{\rm t}$ separately. Then, the combination of these four contributions has been summed in quadrature (column 6), thus providing a conservative estimate of the error bar, which is  probably slightly overestimated since the covariance between
the stellar parameters has been neglected.
However, one should expect that the errors in the [X/H] abundance differ from those of the [X/Fe] ratios, since in many cases the effect of changing the  stellar parameters is similar for lines of different elements. Thus, the [X/Fe] ratio is often less sensitive to stellar parameter uncertainties. The error budget in the [X/Fe] scale therefore is reported in col.7., confirming that $ \sigma_{\rm total([X/H])} $ is larger than $ \sigma_{\rm total([X/Fe])} $ for every species.

\end{enumerate}

\begin{table}
\caption{Mean line-to-line scatter when two or more lines are available.
}
\begin{tabular}{lcccccc}
\hline\hline
Line & $\langle \sigma \rangle$ & $\sigma_{\rm{max}}^*$ & $N_{max}^{**}$ \\
\hline

${\rm   [\ion{Mg}{I}/H] }$&     0.02    &       0.05    & 8\\
${\rm   [\ion{Mn}{I}/H] }$&     0.03    &       0.05    & 14\\
${\rm   [\ion{Fe}{I}/H] }$&     0.04    &       0.07    & 139\\
${\rm   [\ion{Fe}{II}/H]}$&     0.02    &       0.07    & 11\\
${\rm   [\ion{Ni}{I}/H] }$&     0.03    &       0.05    & 30\\
${\rm   [\ion{Cu}{I}/H] }$&     0.03    &       0.07    & 6\\
${\rm   [\ion{Zn}{I}/H] }$&     0.02    &       0.07    & 3\\

\hline
 \label{tab:Scatter}
\end{tabular}

$^*$ Largest line-to-line scatter when two or more lines are available.

$^{**}$ $N_{max}$ is the maximum number of available lines.
\end{table}

\input{tab-sensitivity.tex}

Finally, the total error budget has been estimated by taking into account the errors from stellar parameter uncertainties as well as random errors.
We thus calculated $\sigma_{\rm all([X/Fe])}$ (Table~\ref{tab:Sensitivity}, col.8) for every given element of every given star as\\ \\
$\sigma_{\rm all([X/Fe])} = \sqrt{\sigma^2_{\rm total([X/Fe])}+\left(\frac{\sigma_{\rm N}}{\sqrt{N}}\right)^2   } $,\\ \\
where $\sigma_{\rm N}$ is the line-to-line scatter and $N$ is the number of analysed atomic lines. If the number of lines is $N=1$, we adopted $\langle \sigma \rangle $ from Table~\ref{tab:Scatter} for a given element as $\sigma_{\rm N}$. The typical errors (in [X/Fe]) for the abundances are between 0.03 and 0.12~dex, but most of these errors are smaller than 0.07~dex.

\subsection{The final catalogue}
\label{sec:catalog}
\begin{figure*}[htb]
   \advance\leftskip 0cm
   \includegraphics[scale=0.25]{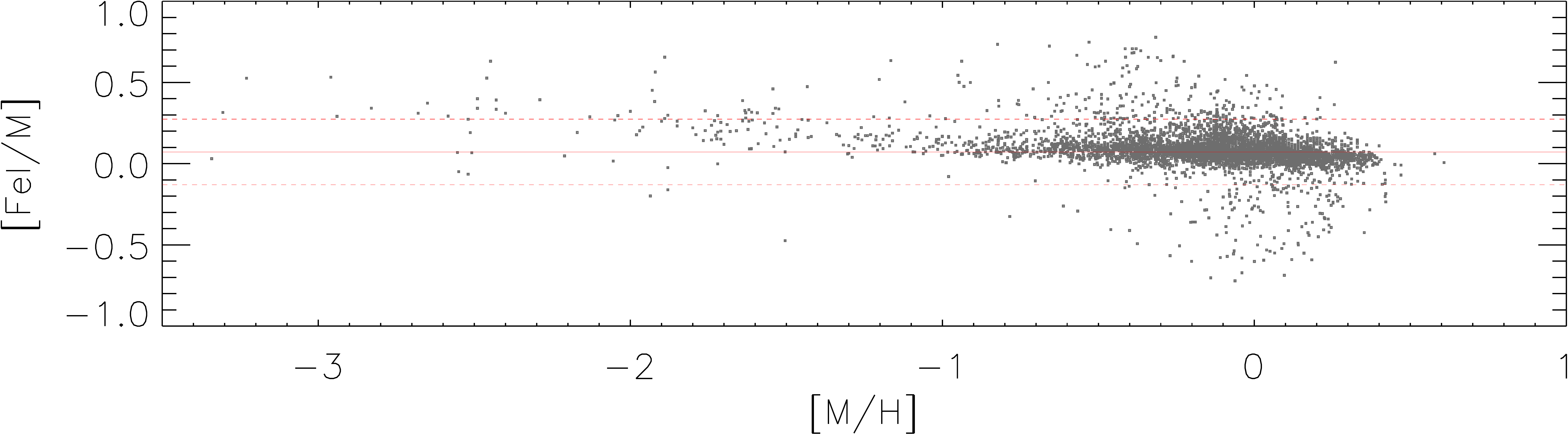}
  \caption{
Comparison between the [\ion{Fe}{I}/H] and [M/H] (i.e. [\ion{Fe}{I}/M] ratio) vs. the metallicity for the whole initial sample of
stars. The mean of [\ion{Fe}{I}/M] is indicated by a solid red line. Dashed lines indicate the difference of $\pm2\sigma$
that we adopted to define the final sample.
  }
  \label{fig:FE1_metalicity}
  \end{figure*}

The metallicity value ([M/H]) derived by AMBRE corresponds to the mean metallicity of a star. However, we also 
derived the \ion{Fe}{I} and \ion{Fe}{II} abundances  in the present work.  
Since this mean metallicity and these iron abundances were computed in different and slightly independent ways, 
it is important to check whether they are consistent between each other.
We thus provide this comparison in Fig.~\ref{fig:FE1_metalicity}. First, the bias of [\ion{Fe}{I}/H] according to [M/H] is +0.072~dex with a scatter $\sigma=0.10$~dex. For metal-poor stars, we detect slightly higher \ion{Fe}{I} abundances than for metal-rich stars. We did not detect any other notable systematic effects (regarding for instance some dependencies with stellar parameters).

    \begin{figure*}[htb]
   \advance\leftskip 0cm
   \includegraphics[scale=0.25]{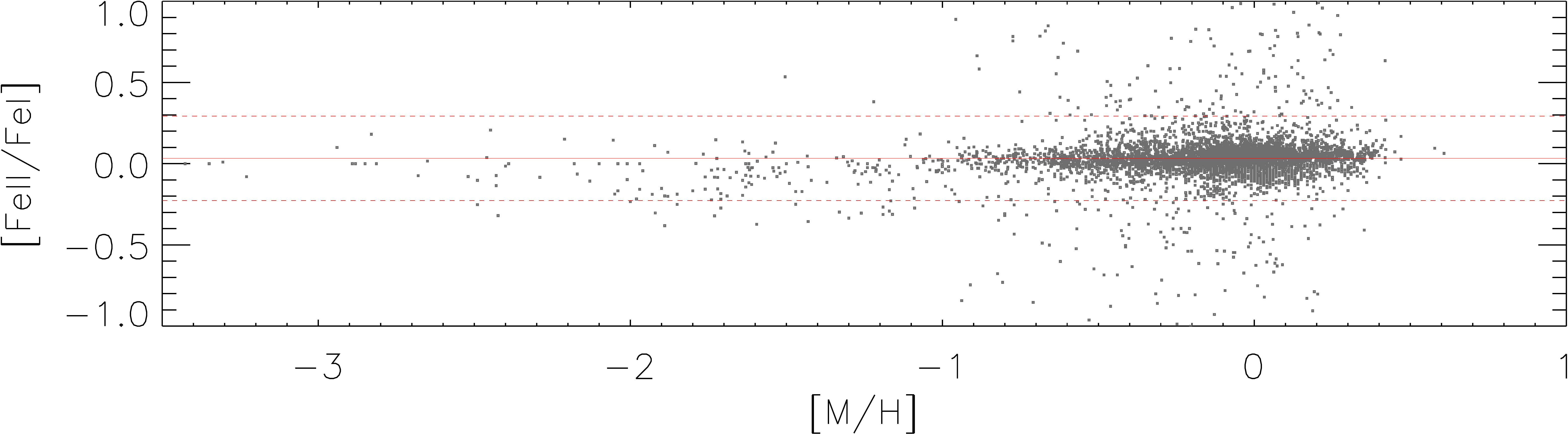}
  \caption{
Same as Fig.~\ref{fig:FE1_metalicity} but for the \ion{Fe}{II} and \ion{Fe}{I} abundances.
  }
  \label{fig:FE1-FE2}
  \end{figure*}

Secondly, Figure~\ref{fig:FE1-FE2} compares the neutral and ionised iron abundances to check the ionisation equilibrium. We detect a very small bias (0.033~dex) for [\ion{Fe}{II}/\ion{Fe}{I}] and a scatter $\sigma=0.13$~dex. One can see that ionisation equilibrium is better preserved for metal-rich stars (for which the NLTE effects are believed to be smaller). For metal-poor stars ([M/H]$<$-1.0~dex) the [\ion{Fe}{II}/\ion{Fe}{I}] values are lower by about $0.101$~dex. Obviously, NLTE effects are significantly stronger for metal-poor stars, but the prime effect is over-ionisation. However, our metal-poor stars are mostly giants ($T_{\rm eff} < 5250~K$, ${\rm log}~g < 3.5$) for which the gravity is derived with larger uncertainties. Anyway, the detected small, unbalanced ionisation equilibrium and stronger drift of [\ion{Fe}{I}/H] from [M/H] could be caused by some NLTE effects and larger uncertainties of atmospheric parameters, especially for these metal-poor stars. 

On the other hand,
there are some stars that depart in [\ion{Fe}{I}/M] by more than $2\sigma$ with respect to the mean [\ion{Fe}{I}/M] at a given metallicity. Also there are some stars for which the ionisation balance is not preserved ($|$[\ion{Fe}{II}/\ion{Fe}{I}]$-\langle$[\ion{Fe}{II}/\ion{Fe}{I}]$\rangle| > 2\sigma$). To\ avoid any spurious effect on abundances, we decided that, if at least one of these two conditions is not preserved for a given star, it is rejected from further analysis and from the final catalogue.

 \begin{figure}[!htb]
   \advance\leftskip 0cm
   \includegraphics[scale=0.44]{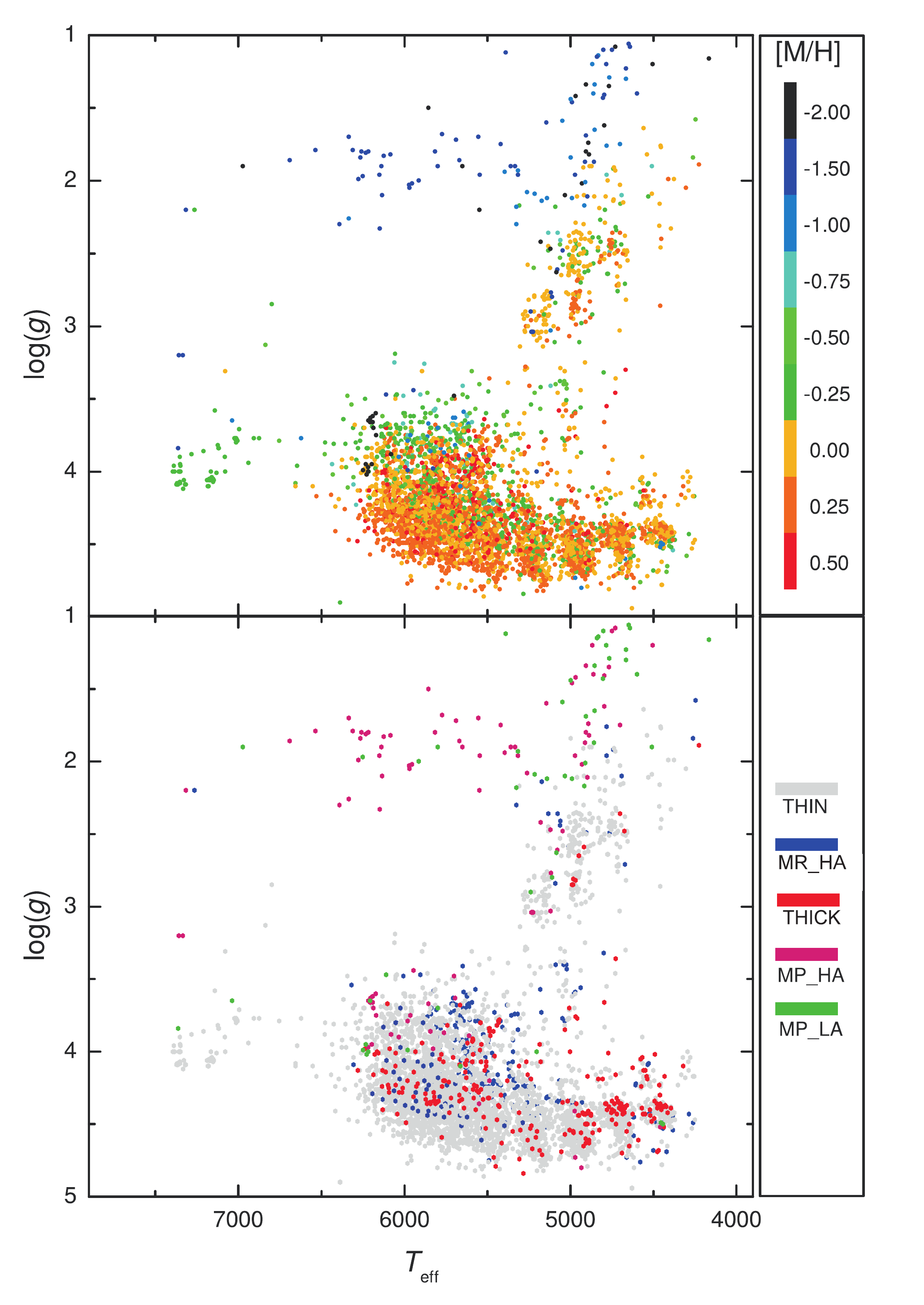}
  \caption{Effective temperature -- surface gravity diagram of the main sample stars. The metallicity is coded in colours in the upper panel and the assigned Galactic component is colour coded in the lower panel (tags in the legend are the same as in Table~\ref{tab:all_stars}).
  }
  \label{fig:hr}
  \end{figure}

As a consequence, we omitted 987 stars from the primary list of 5,960
because of our $v{\rm sin}i$ criteria and 307 additional stars because of 
the M-FeI-FeII test. The cool portion of the
main sequence ($T_{\rm eff} < 4500~K$ and ${\rm log}~g < 4.25$) shows 
slightly lower gravity than expected (${\rm log}~g > 4.3$) as already pointed out by several independent analyses (e.g. within the Gaia-ESO survey). All stars (62) that show very large (more than 0.5~dex) positive or negative [\ion{Fe}{I}/M] and [\ion{Fe}{II}/\ion{Fe}{I}] belong to that cool main-sequence potion. 
As already pointed out by \citet{Worley2012}, one of the (several) causes of this problem could be the normalisation procedure that
may not be as robust for these cool stars. 
We also checked whether this issue could affect the individual chemical
abundances, and we confirmed that the [\ion{X}{I}/Fe] ratios are not very sensitive to the ${\rm log}~g$ for these stars. Such stars have been however rejected from
the initial sample. 

Finally, the AMBRE catalogue of iron-peak elements contains 4,666 stars and at least Fe and Mg are provided for all of these stars. The Mn, Ni, Cu, and Zn lines were too weak in some metal-poor stars, thus Mn, Ni, Cu, and Zn were provided for stars 4646, 4643, 4602, and 4646, respectively. We point out again that because of the lack of NLTE correction tables available for all chemical species studied, all forthcoming discussions are based on the derived LTE abundances.
This set is called the \textit{main} sample hereafter. The effective temperature versus gravity  diagram  of this sample is shown in Fig.~\ref{fig:hr}. 
It consists of 4\,252 dwarf stars (${\rm log}~g \ge 3.5$) and 414 giant stars (${\rm log}~g < 3.5$). Among these 4,666 stars, 783 are metal poor ([M/H]$<$-0.5) and 144 are very metal poor ([M/H]$<$-1.0). The final catalogue of stellar photospheric elemental abundances is listed in Table~\ref{tab:all_stars}.

   \begin{table*}
\begin{center}
      \caption{Chemical abundances of iron-peak elements and magnesium in the solar neighbourhood.
}
          \label{tab:all_stars}
      \[
      \resizebox{\textwidth}{!}{%
         \begin{tabular}{llrcccccccccccccccccccc}
            \hline
            \noalign{\smallskip}
            RA & DEC & Sepctrograph 
            & [\ion{Mg}{I}] & $ \sigma_{\rm all\left[\frac{MgI}{FeI}\right]} $  
            & [\ion{Mn}{I}] & $ \sigma_{\rm all\left[\frac{MnI}{FeI}\right]} $  
            & [\ion{Fe}{I}] & $ \sigma_{\rm all\left[\frac{FeI}{H}\right]} $  
            & [\ion{Fe}{II}] & $ \sigma_{\rm all\left[\frac{FeII}{H}\right]} $  
            & [\ion{Ni}{I}] & $ \sigma_{\rm all\left[\frac{NiI}{FeI}\right]} $  
            & [\ion{Cu}{I}] & $ \sigma_{\rm all\left[\frac{CuI}{FeI}\right]} $  
            & [\ion{Zn}{I}] & $ \sigma_{\rm all\left[\frac{ZnI}{FeI}\right]} $  
            & GC*  

             \\ 
            \noalign{\smallskip}
            \hline
            \noalign{\smallskip}
11.11568        &       -65.65181       & FEROS &   0.08 & 0.07&   0.07 & 0.06&   0.09 & 0.05&   0.06 & 0.05&   0.01 & 0.07&   0.02 & 0.05&   0.08 & 0.05 & THIN   \\
        \end{tabular}}
      \]
  \end{center}
Notes. This is a small part of the table (The full table is available at the CDS).\\
$^*$Galactic components (GC) defined in Sect.~\ref{sec:definition_of_components} are listed: thin disc (THIN), thick disc (THICK), metal-rich high-$\alpha$ sequence (MR\_HA), metal-poor low-$\alpha$ sequence (MP\_LA) and metal-poor high-$\alpha$ sequence (MP\_HA).
\end{table*}

\section{Chemical definition of the Galactic components}
\label{sec:definition_of_components}

In this section, we describe the procedure adopted to define 
chemically the different Galactic components that are explored later in this article. 

  \begin{figure*}[htb]
  \centering
   \advance\leftskip 0cm
   \includegraphics[scale=0.27]{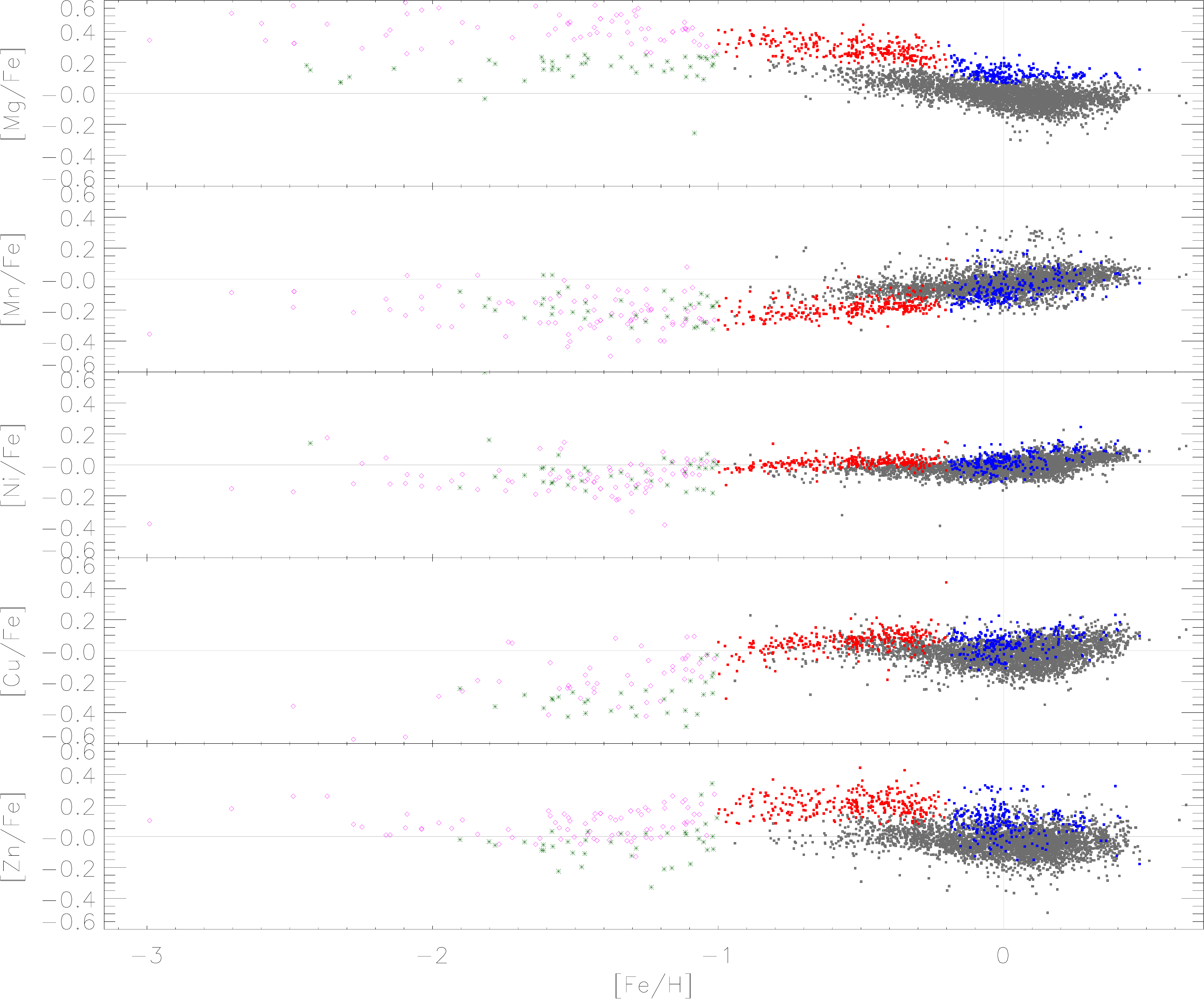}
  \caption{
Observed element vs. iron abundance ratios as a function of the iron abundance. The different
Galactic components defined in Sect.~\ref{sec:definition_of_components} are plotted: thin disc (grey dots), thick disc (red dots), metal-rich high-$\alpha$ sequence (blue dots), metal-poor low-$\alpha$ sequence (green asterisk), and metal-poor high-$\alpha$ sequence (magenta rhombus); the light grey lines represent the solar values.
  }
  \label{fig:ELEMENTS_separation}
  \end{figure*}

The abundances of the $\alpha$ elements are a well-known chemical probe for chemically defining thin and thick discs (see \citealt{Fuhrmann1998, Recio2014, Mikolaitis2014, Rojas2016}). Thin disc stars are known to have significantly lower [$\alpha$/Fe] abundances than thick disc stars, making both populations easy to disentangle chemically. In \citet{Mikolaitis2014}, we already discussed that [Mg/Fe] is one of the best $\alpha$ elements to define the Galactic disc components. 
We therefore adopt the same procedure in the present study, and the top panel of Fig.~\ref{fig:ELEMENTS_separation} shows the [Mg/Fe] distribution as a function of [Fe/H] for our final sample with a specific colour code for the different Galactic components defined below. 
This definition is based on [Mg/Fe] distributions in various metallicity bins, where the sample stars are tagged as thin ($\alpha$-rich) or thick ($\alpha$-poor) disc members.

As expected, the stars with [Fe/H]$>$-1.0~dex clearly separate into two 
distinct $\alpha$-poor and $\alpha$-rich (thin and thick disc) populations. 
One can also see that the thick disc population has a gap in metallicity at about [Fe/H]$\approx$-0.2~dex. This gap lies at the same location as in \citet{Adibekyan2011,Adibekyan2012}, who used it to define the thick disc and $\alpha$-rich metal-rich sequence. 
In the following, we adopted this separation at [Fe/H]$\approx$-0.2~dex to define these
two subpopulations of the disc. 

As for the metal-poor stars ([Fe/H]$<$-1.0~dex), the top panel of Fig.~\ref{fig:ELEMENTS_separation} shows that this subpopulation is also very well separated into $\alpha$-rich and $\alpha$-poor components. This trend was already observed by \citet{Nissen2011}.
We identified the low-$\alpha$ and high-$\alpha$ metal-poor sequences using the same separation function as in \citet{Nissen2010, Nissen2011} for metallicities
in the range -1.6~dex~to~-1.0~dex. For [Fe/H]$<$-1.6~dex range, we adopted 
a separation criterium at [Mg/Fe]=0.25~dex to define the low- and high-$\alpha$ sequences. These authors have shown that low-$\alpha$ and high-$\alpha$ metal-poor sequences separate visually in the [Fe/H] versus [$\alpha$/Fe] plane. They also have checked kinematically, via a Toomre
diagram, that the two sequences are distinct.
We point out that the metal-poor high-$\alpha$ group could contain both halo and thick disc stars because these populations are chemically indistinguishable (see e.g. \citealt{Nissen2010}).

    \begin{figure*}[htb]
      \centering
   \advance\leftskip 0cm
   \includegraphics[scale=0.38]{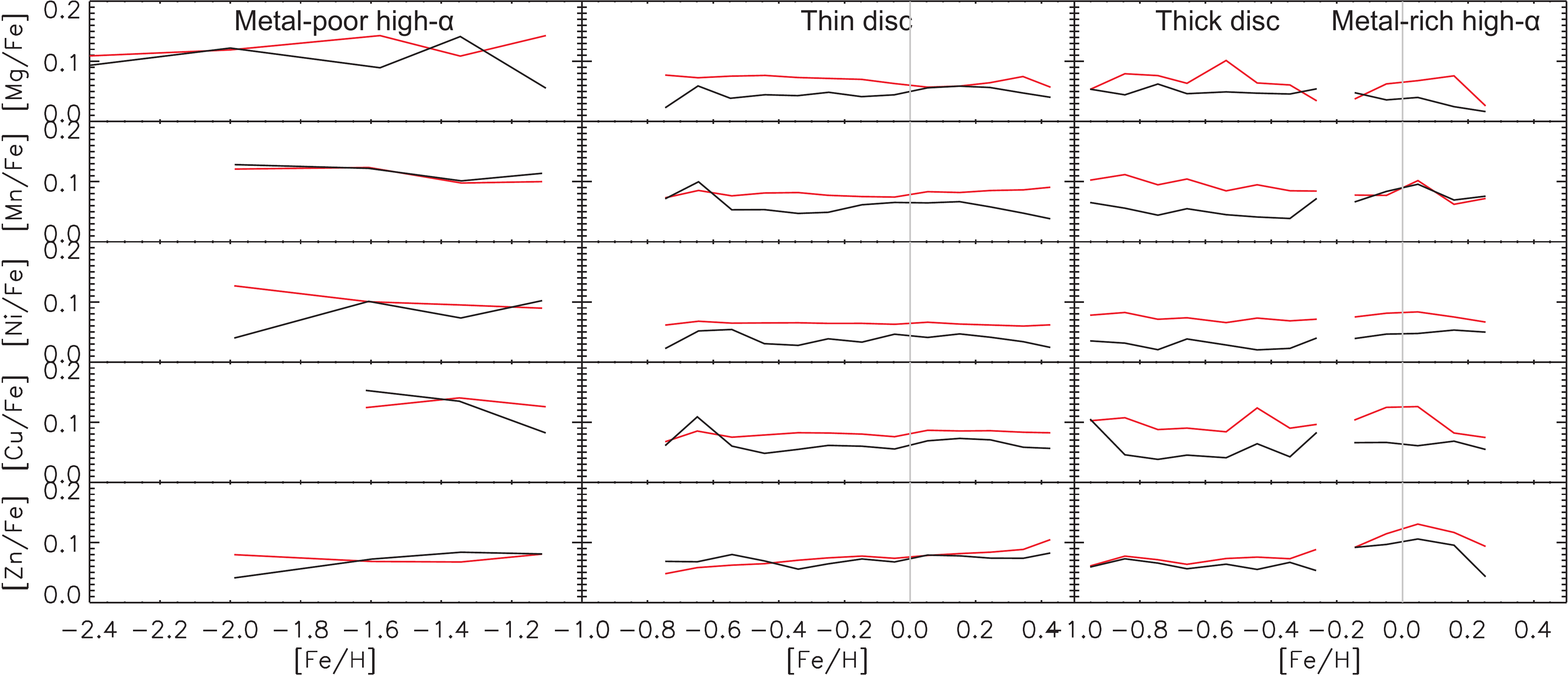}
  \caption{
Mean expected errors (red line) and measured dispersion (black line) for the [El/Fe] ratios with respect to the metallicity (binned every 0.2~dex and 0.3~dex for the metal-rich and metal-poor samples, respectively) for the main Galactic components. 
  }
  \label{fig:EL-scatter}
  \end{figure*}

If these five Galactic components are mixed well, chemically homogeneous, and well defined, one could expect that the scatter inside each component should be similar to the expected measurement errors. If not, either one of the component is not very well
defined and/or the measurement errors are underestimated; for example, some
of the adopted atomic lines could be sensitive to NLTE effects. 
As a sanity check, we thus show a comparison between expected errors (red line) and measured dispersion (black line) in metallicity bins in the top panel of Fig.~\ref{fig:EL-scatter}. We require every metallicity bin to contain at least five stars. Otherwise the statistical result is not trustworthy, so it is not shown in the plot. That is why the metal-poor part of Fig.~\ref{fig:EL-scatter} contains different number of data points for different elements. It can be seen that the observed dispersion around the trends is fully described by the expected errors, leaving little
room for astrophysical dispersion. 
However, we point out that errors are overestimated in some metallicity bins. We indeed already noticed that the atmospheric parameters errors that we use are too pessimistic for many stars in our sample. We adopted the general typical errors (see footnote in Table~\ref{tab:Montecarlo}) provided by AMBRE (neglecting the co-variance and hence overestimating the errors) and summed than quadratically to compute the total error.  For  metal-poor stars, only the high-$\alpha$ sequence is shown because the low-$\alpha$ is too weakly populated.

Moreover, Fig.~\ref{fig:EL-scatter} indicates that the scatter in the thin and thick disc sequences is very low ($\approx$0.05 dex). Actually, it provides a very important observational constrain for Galactic evolutionary models that study the efficiency of the possible stellar radial migration. These models should satisfy the observed upper limit of dispersions for both discs.

In summary, we therefore defined five different components in our sample data:
(1) the thin disc ($\alpha$-poor) population that covers the metallicity range between [Fe/H]$\approx$-0.8~dex to [Fe/H]$\approx$+0.5~dex and consists of 3,949 stars (84.5\% of the total sample); (2) the metal-rich high-$\alpha$ sequence from [Fe/H]$\approx$-0.2~dex to [Fe/H]$\approx$~+0.3~dex (260 stars, 6\% of the total sample); (3) the metal-poor ($\alpha$-rich) thick disc that starts from [Fe/H]$\approx$-1.0~dex and ends at [Fe/H]$\approx$-0.2~dex (313 stars, 6.5\% of the total sample). There are also   144 metal-poor ([Fe/H]$<$-1.0~dex) stars (3\% of the total sample): (4) 49 of which are 
low-$\alpha$ stars, whereas (5) 95 of which are high-$\alpha$ stars. We have a rather similar population ratios between these five Galactic components as \citealt{Adibekyan2012}, but our final sample is more than four times larger. Every star from our final catalogue is tagged as a member of one of these five components in last column of Table~\ref{tab:all_stars}. The assigned Galactic components are colour coded in lower panel of effective temperature -- surface gravity diagram (Fig.~\ref{fig:hr}).

\section{Properties of the iron-peak elements in the different Galactic disc components}
\label{sec:IroninSolarN}
\begin{figure*}[htb]
   \advance\leftskip 0cm
   \includegraphics[scale=0.50]{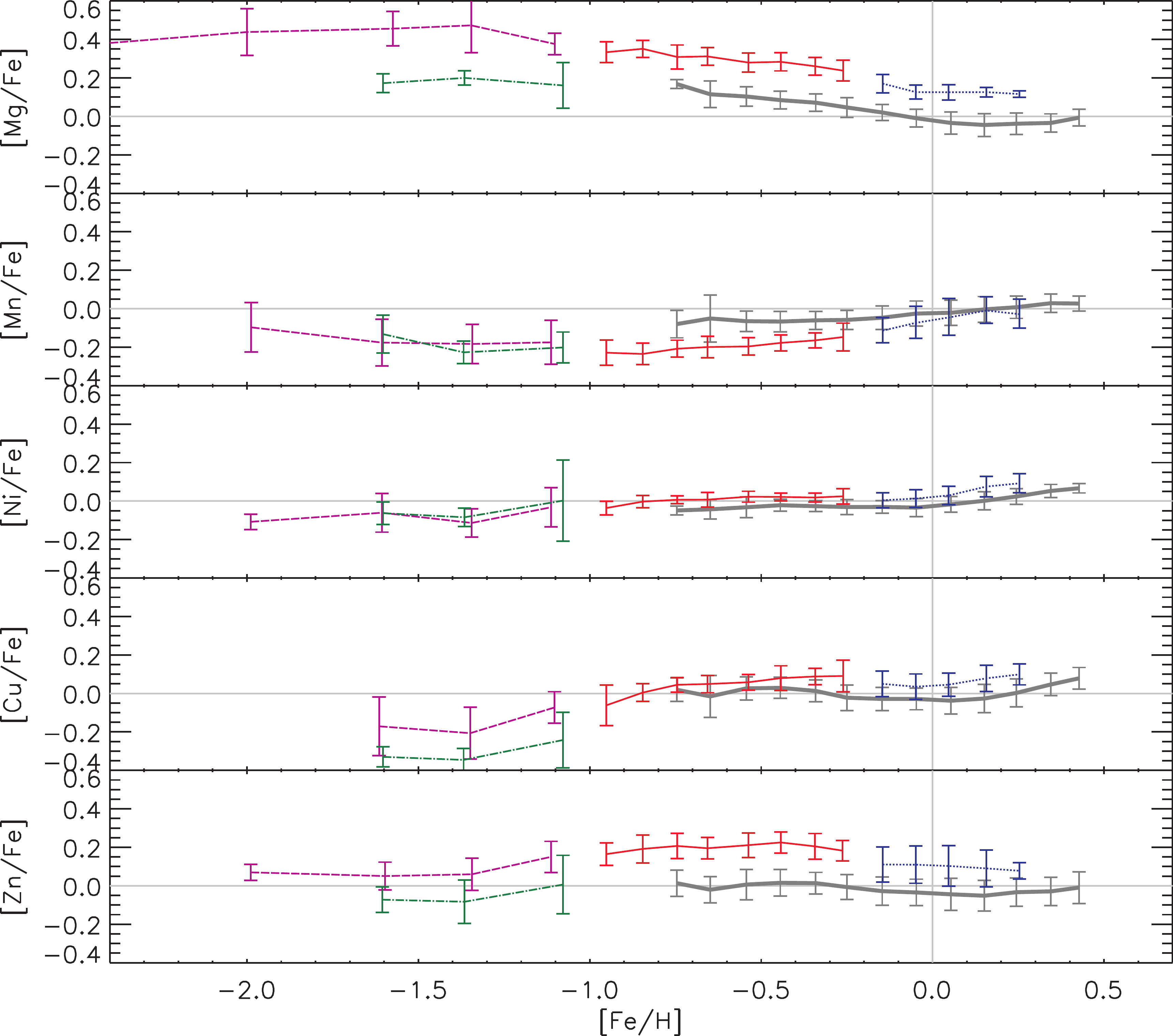}
  \caption{Same plot as Fig.~\ref{fig:ELEMENTS_separation}, but data are averaged in [Fe/H] bins. Thin disc (thick grey solid line), thick disc (red solid line), metal-rich high-$\alpha$ sequence (blue dotted line), metal-poor low-$\alpha$ sequence (dash-dot green line), and metal-poor high-$\alpha$ sequence (dashed magenta line); the light grey lines represent the solar values.
    The binning structure is the same as in Fig~\ref{fig:EL-scatter}. The error bars represent the standard deviation associated with the mean value. 
}
  \label{fig:EL-trends}
  \end{figure*}

\begin{figure*}[htb]
   \advance\leftskip 0cm
   \includegraphics[scale=0.50]{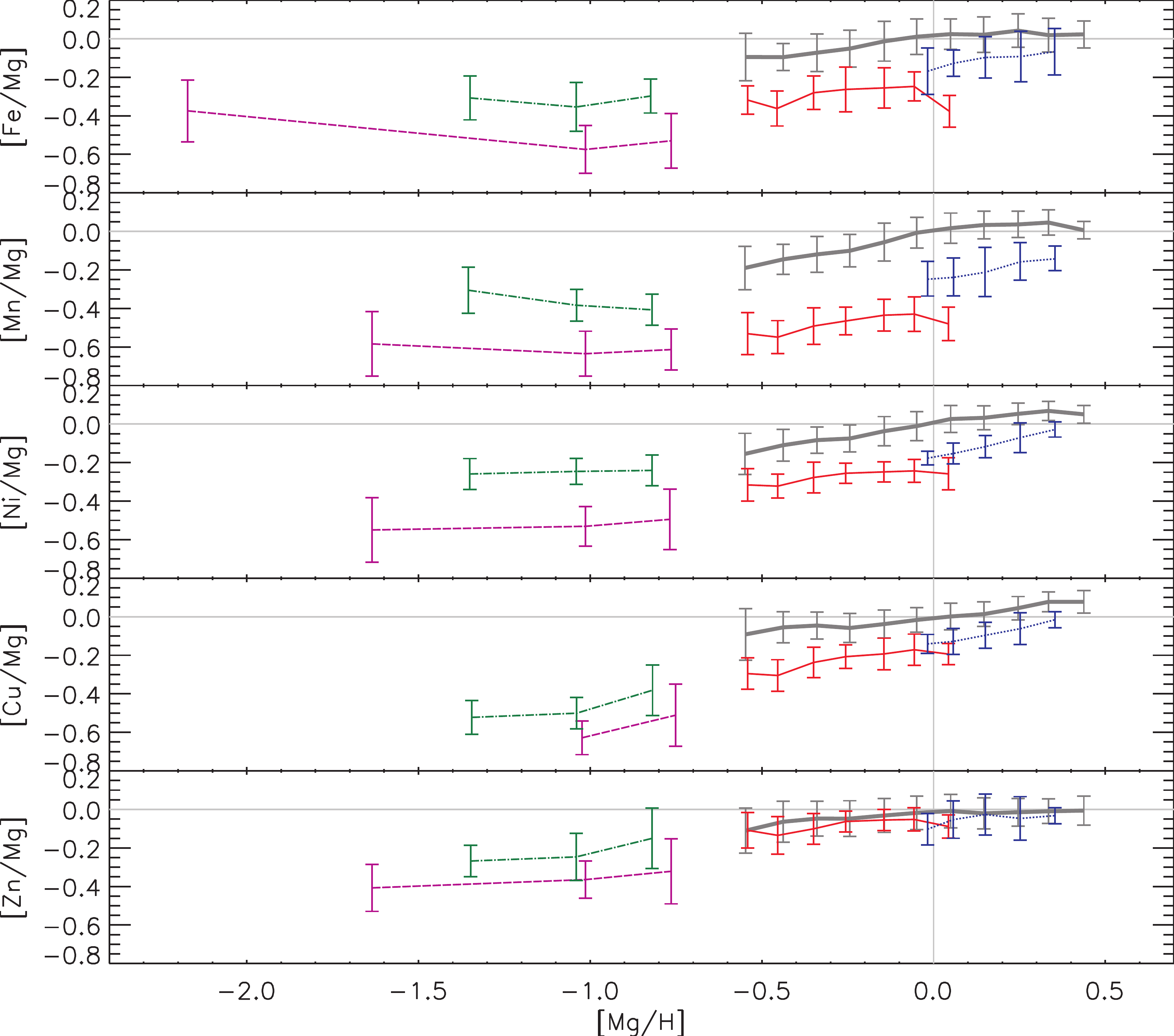}
  \caption{Same plot as Fig.~\ref{fig:EL-trends}, but for the [X/Mg] vs. [Mg/H] ratios. The binning structure is the same as in Figs~\ref{fig:EL-scatter}~and~\ref{fig:EL-trends}. The error bars represent the standard deviation associated with the mean value.
}
  \label{fig:EL-trends-mg}
  \end{figure*}

In this section, we present and discuss the iron-peak abundances for stars of our sample (4,666~stars). We mostly focus on the disc components, although we further compare our results to other Galactic components.  First we present the iron-peak chemical characterisation of the defined Galactic component (Figs.~\ref{fig:EL-trends} and \ref{fig:EL-trends-mg} which show the mean tendencies), and then we discuss them in comparison  with previous high-resolution spectroscopic studies devoted to the chemical properties of the Galactic disc.

 \subsection{Manganese}
 \label{sec:Manganese}
 
The only stable manganese isotope $^{55}$Mn is produced by type-Ia and type-II supernovae
in which thermonuclear explosive silicon burning and alpha-rich freeze-out via synthesis of $^{55}$Co are active. 
They decay then via $^{55}$Fe isotope to the stable $^{55}$Mn (\citealt{Truran1967}). However, the yields from type-Ia supernovae are much larger (approximately by two orders of magnitude; see \citealt{Nomoto1997b,Nomoto1997,Kobayashi2006}), so manganese is logically considered a characteristic yield of type-Ia supernovae 
(\citealt{Seitenzahl2013}).

In our sample stars, we observe an increasing trend of [Mn/Fe] with [Fe/H] (see Fig.~\ref{fig:ELEMENTS_separation} and \ref{fig:EL-trends}).
Moreover, thin and thick disc populations are seen to follow the same trend with metallicity. However, [Mn/Fe] abundances of thick disc stars tend to be smaller than thin disc 
stars by about 0.05~dex in the same metallicity range (Fig.~\ref{fig:EL-trends}). 
More metal-rich high-$\alpha$ sequence stars could also be slightly less enriched in manganese than thin disc stars.
Finally, the high-$\alpha$ and low-$\alpha$ metal-poor populations overlap and do not have any distinguishable trends, although they seem to follow the thick disc trend forming a plateau around [Mn/Fe]=-0.25~dex. 

The number of published observations of manganese is very large,
but only three studies \citep{Adibekyan2012,Battistini2015,Hawkins2015} report statistically representative samples (1\,111, 714, 3\,200 stars, respectively).
The manganese abundances of the thin and thick disc was first
explicitly studied by \citet{Feltzing2007} who kinematically
divided their sample of 95 dwarf stars into thin and thick disc components. 
Their thick disc sample is found to be lower in [Mn/Fe] than their thin disc stars by about 0.05~dex at the same metallicities, which agrees with our results based on a much more statistically significant sample spanning a broader range of metallicity.
\citet{Adibekyan2012} also clearly showed  an increasing trend of [Mn/Fe] versus [Fe/H] for their sample of 1111 HARPS stars, which were chemically separated thin and thick disc samples in a similar manner as in our study. 
They concluded that there are some hints indicating that the two discs have different [Mn/Fe] ratios, but there is no clear boundary between
both discs. On the contrary and owing to our smaller dispersion, such a thin/thick disc dichotomy appears in a much clearer way in our sample, particularly for metallicities that are smaller than [Fe/H]$\sim -0.2$~dex (see, for instance, 
Fig.~\ref{fig:EL-trends}). 
Furthermore, in their smaller sample \citet{Battistini2015} found no
significant distinction between thin and thick disc populations contrary
to what we clearly see. This could be because of a larger dispersion in their chemical analysis.
However, NLTE corrections to their Mn abundances revealed
a much clearer distinction.
Finally, \citet{Hawkins2015} reported high accuracy measurements of manganese abundance from the APOGEE survey for 3200 giant stars. They found different behaviours of manganese in the thin and thick discs: [Mn/Fe] abundance ratios
in thin disc stars are generally larger than in thick disc stars. This is in 
good agreement with our results. For example, their thin and thick disc trends in [Fe/H] varying from -0.7~dex to -0.5~dex differ by 0.09-0.06 dex, which is very close to what we see in our data.

On the other hand, our metal-poor sample (144 stars) can be compared to the 
smaller samples of \citet{Reddy2006} and \citeauthor{Reddy2008} (\citeyear{Reddy2008}; 36 stars), \citeauthor{Sobeck2006} (\citeyear{Sobeck2006}; 200 globular cluster stars), \citeauthor{Nissen2011} (\citeyear{Nissen2011}; 98 stars), and \citeauthor{Ishigaki2013} (\citeyear{Ishigaki2013}; 78 stars).
These investigations, in agreement with our results, observed an almost flat trend
at [Mn/Fe]$\sim$-0.3~dex from [Fe/H]$\approx$-1.6~dex to -1.0~dex . 
We finally point out that \citet{Nissen2011} and \citet{Ishigaki2013} also performed a chemical separation between high-$\alpha$ and low-$\alpha$ metal-poor stars.  Similar to our results, based on a sample that was two times larger, they did not find any distinguishable [Mn/Fe] separation between the high-$\alpha$ and low-$\alpha$ metal-poor populations. 

We have however several results for [Fe/H]$<$-2.5~dex that correspond to extremely metal-poor (EMP) stars. The number of such stars is too small to have any statistics, but we can compare tendencies of our results with largest literature abundance sources of EMP stars; these sources are \citet{Cayrel2004} (70 stars) and \citet{Lai2008} (28 stars). The [Mn/Fe] behaviours from both literature sources scatter by 0.3~dex around [Mn/Fe]$\approx$-0.4~dex and our points are in agreement with these literature trends.

It is interesting to compare our results with other Galactic substructures, even if there is no bulge or any dwarf spheroidal galaxy (dSph) stars in our sample. Mn abundances of bulge giants from \citet{Barbuy2013} (56 bulge giant stars) show a good agreement with our thin and thick disc stars for metallicities higher than [Fe/H]$>\approx$-0.7~dex, whereas, for lower abundances, the [Mn/Fe] values in the bulge appear to decrease faster than for the thick disc stars with [Fe/H]$>\approx$-0.7~dex. Similarly, [Mn/Fe] abundances in dSphs are approximately at a similar level as our thin and thick disc stars (\citealt{North2012, Jablonka2015, Romano2011, Sbordone2007}). However, some of them show very specific trends that are opposite from what we see in the solar neighbourhood. For example, [Mn/Fe] in the dSph Sculptor galaxy (\citealt{North2012}) slowly decreases from -0.2~to~-0.5~dex with increasing metallicity (from~-1.8~to~-0.8~dex) and also $\omega$~Cen exhibits a decrease from -0.2~to~-0.7~dex with increasing metallicity (from~-1.8~to~-0.8~dex). This very specific behaviour of [Mn/Fe] reveals that the chemical enrichment history is different from that of Milky Way.

From the point of view of Galactic chemical evolution (GCE), it would be interesting to use reference elements other than iron because its nucleosynthesis channels are not uniquely coming both from explosive nucleosynthesis in type-II or type-I supernovae. A good choice would be oxygen (\citealt{Cayrel2004}) since it is the most abundant element after H and He and it comes from a single source. However, because of observational difficulties, it would considerably degrade the accuracy of the derived trends. Another natural choice is magnesium, since its abundances are accurately determined and it is also mainly formed in massive supernovae (\citealt{Shigeyama1998}). In the following, we thus discuss our iron-peak abundances with respect to the Mg abundances (see Fig.~\ref{fig:EL-trends-mg}).

In our sample stars, we observe a global underabundance of [Mn/Mg] for the metal-poor stars.
This [Mn/Mg] ratio then increases with [Mg/H] after type-I supernovae start to enrich the interstellar medium
(ISM; see Fig.~\ref{fig:EL-trends-mg}).
This is expected
since type-II supernovae produce
significantly less manganese than magnesium (e.g. yields by \citealt{Kobayashi2006}).
We also point out that the [Mn/Mg] ratio in high-$\alpha$ metal-poor stars is much larger than
in low-$\alpha$ metal-poor stars
(by $\approx$-0.2~dex). However, this is
probably caused by magnesium, since both metal-poor trends also differ by the same amount
in the [Mn/Fe] versus [Mg/Fe] plot (see Fig.~\ref{fig:EL-trends-mg}, top panel).
Moreover, the thin and thick disc trends are seen to be much more distinct in [Mn/Mg] (by
$\approx$0.3~dex) than in [Mn/Fe] plane of Fig.~\ref{fig:EL-trends}. This manganese-magnesium behaviour suggests that the manganese enrichment history clearly has the opposite behaviour.

Actually, there are rather few studies in the literature devoted to the
[Mn/$\alpha$]-[$\alpha$/H]. The only valuable sources for comparison are
\citet{Battistini2015}, which is devoted to rather metal-rich stars, and \citet{Hawkins2015}. 
\citet{Battistini2015} used  another $\alpha$ element (titanium) as a reference element and showed fully separated
[Mn/Ti]-[Ti/H] trends (by $\approx$0.2~dex) for the thin and thick discs.The [Mn/Ti]
ratios for both discs are closer than the  [Mn/Mg] discs
probably because the yields of Ti from type-I supernovae
are larger than those of Mg (e.g. yields by \citealt{Kobayashi2006}).
Nevertheless the trends of \citet{Battistini2015} are in general agreement with our results.
\citet{Hawkins2015} suggested a new set of chemical abundance planes: [Mn/Mg],
[$\alpha$/Fe], [Al/Fe], and [C+N/Fe], which can help to disentangle Galactic components in a
clean and efficient way independent of any kinematical data.
From their Fig.~9, we actually see that the
[Mn/Mg] ratio plays a key role in confirming the thin and thick disc separation.
Of course, that method should be studied in much more detail.
However, we agree with \citet{Hawkins2015} that
the Mn-$\alpha$ ratio could be another good tool to chemically disentangle thin and thick disc stars.

In Fig.~\ref{fig:EL-trends-mn}, we show some interesting results about the [Fe,Ni,Cu, and Zn/Mn] trends with respect to [Mn/H]. The thick disc [Fe,Ni,Cu, and Zn/Mn] trends are generally found to be higher than the thin disc trends. The largest difference is found for [Zn/Mn] where the thin and thick discs differ by about $\approx$0.25~dex. In the metal-poor regime, the metal-poor trends of [Fe,Ni/Mn] overlap and are close to $\approx$0.20~dex. This is not the case for the [Cu,Zn/Mn] trends in the metal-poor regime for which the high-$\alpha$
sequence is found to be $\approx$0.20~dex larger than the low-$\alpha$ sequence.

\subsection{Nickel}
 \label{sec:Nickel} 

Nickel is also produced by silicon burning, but the [Ni/Fe] ratio in yields of 
type-Ia and type-II supernovae is similar (see \citealt{Nomoto2013}). 
This leads to the characteristic flat [Ni/Fe] versus [Fe/H] behaviour
shown in the second panel of Fig.~\ref{fig:ELEMENTS_separation}. The metal-poor stars are found in a plateau around [Ni/Fe]$\approx$-0.1~dex with a small scatter.
However, metal-poor trends are exactly equal between high-$\alpha$ and low-$\alpha$ stars. 
The thick disc population forms a plateau around [Ni/Fe]$\approx$0.0~dex with a very small scatter ($\sigma$=0.04~dex). This plateau is also seen for the metal-rich high-$\alpha$ sequence with a soft increase of [Ni/Fe] at supersolar metallicities. 
The [Ni/Fe] distribution of the thin disc is very tight and steady around [Ni/Fe]$\approx$0.0~dex with a small increase in the supersolar metallicity regime as well. Generally thin and thick disc populations overlap with no strong evidence of any chemical separation, although thin disc stars could reveal a slightly smaller [Ni/Fe] ratio.
 
 To date, the largest samples of observed [Ni/Fe] were studied by \citet{Bensby2014} (714 stars), \citet{Adibekyan2012} (1111 stars), and also \citet{Hawkins2015} (3200 APOGEE giant stars). Regardless the adopted method for the thin/thick disc separation (\citealt[][kinematically]{Bensby2014}; \citealt[][chemically and kinematically]{Adibekyan2012}; \citealt[][chemically]{Hawkins2015}), 
the chemical patterns of [Ni/Fe] for the thin and thick discs are rather similar to ours.
The [Ni/Fe] plateau was also observed for metal-poor stars (around$\sim$-0.1~dex for [Fe/H]$<$-1.0~dex) by \citet{Reddy2006, Reddy2008} and it agrees with our results. \citet{Nissen2010} found that the high-$\alpha$ sequence is tight and follows the solar Ni value whereas the low-$\alpha$ sequence has a lower [Ni/Fe] ratio by -0.15 -- -0.1~dex. 
However, this finding is not confirmed in a similar study by \citet{Ishigaki2013}
who did not report any different nickel trends between their high-$\alpha$ and low-$\alpha$ samples. Our results support the conclusion of \citet{Ishigaki2013}.

[Ni/Fe] results of EMP stars by \citet{Cayrel2004} (70 stars) and \citet{Lai2008} (28 stars)  scatter by 0.25~dex around [Mn/Fe]$\approx$-0.1~dex. That is in general agreement with our results.

{The Ni abundances in the bulge from \citet{Johnson2011} (156 giant stars) show a similar trend as our thin and thick disc stars, although they seem to be richer by $\approx$0.1~dex in the bulge. Such a small difference could be caused by small systematic offset.

On the other hand, we point out that the trend of [Ni/Mg] with respect
to [Mg/H] (Fig.~\ref{fig:EL-trends-mg}) is very similar to the [Fe/Mg] trend, confirming that nickel and iron are produced in similar supernovae environments. 

We used nickel as a reference element in Fig.~\ref{fig:EL-trends-ni}.
Obviously, since nickel and iron are very close in behaviour, all [Mn,Fe,Cu, and Zn/Ni] ratios with respect to [Ni/H] show very similar trends as was the case for [El/Fe] versus [Fe/H] in Fig.~\ref{fig:EL-trends}.

\subsection{Copper}
 \label{sec:Copper}

Copper is an intermediate element between the iron-peak and the neutron capture elements, thus its nucleosynthesis is very complex. For instance, \citet{Bisterzo2004} highlighted several possible nucleosynthesis scenarios of the Cu origin: (i) explosive nucleosynthesis during type-Ia, (ii) type-II supernovae explosions, (iii) weak \textit{s} process occurring during He-burning in massive stars, (iv) main \textit{s} process during helium burning in low-mass AGB stars, and also (v) weak \textit{sr} process. First, \citet{McWilliam2005}, based on a comparison of Cu in our Galaxy and the Sgr dwarf spheroidal galaxy, showed that the type-Ia supernovae yields of Cu could be negligible and  the type-II supernovae contribution of Cu could be up to 5\%. Thus neutron capture processes are significant contributors of copper in the Galaxy (\citealt{Bisterzo2004}). The Cu contribution of s processes in the solar system is indeed around 27\% (\citealt{Bisterzo2004}) coming from AGB and massive stars. However AGB stars could contribute only marginaly (5\% contribution) since Bisterzo et al. 
claim that the bulk of cosmic Cu has actually been produced by the weak \textit{sr} process.

The Cu abundances derived for our sample stars (see Figures~\ref{fig:ELEMENTS_separation} and \ref{fig:EL-trends})
show that each Galactic component has a specific behaviour in the [Cu/Fe] versus [Fe/H] plane.
First, the thick disc stars look slightly more enriched in Cu than thin disc stars
by about 0.05~dex. The thin disc sample varies around [Cu/Fe]$\approx$0.0~dex with a small positive slope in the supersolar metallicity regime. 
Both discs also show a mild increase in [Cu/Fe] at supersolar metallicities up to [Cu/Fe]$\approx$~0.1~dex. 
On the other hand, the metal-poor stars ([Fe/H$]<$-1.0) show a larger scatter and an increase of [Cu/Fe] towards higher metallicities from [Cu/Fe]$\approx$-0.5 at [Fe/H]$\approx$-2.0 to solar at [Fe/H]$\approx$-1.0, reaching the level of the thick disc stars. High- and low-$\alpha$ sequences partially overlap but low-$\alpha$ stars
are found to have a smaller [Cu/Fe] than the high-$\alpha$ stars by about 0.2~dex.

There are no published observations of very large samples of stellar copper abundances. Homogoenous samples were analysed by \citet{Reddy2008} (60 metal-poor stars), \citet{Reddy2003} (181 metal-rich stars), \citet{Yan2015} (64 late type stars),  and \citet{Mishenina2011} (172 metal-rich stars). The largest non-uniform sample was collected by \citet{Bisterzo2006} (more than 300 stars) from various literature sources.
Generally, our results agrees with \citet{Reddy2008}, where they showed a tight distribution and mild positive trend of a kinematically defined thick disc (up to [Fe/H]=-0.25~dex) sample. Also, similar to our results, the thin and thick disc distributions of \citet{Mishenina2011}, \citet{Reddy2003}, and \citet{Yan2015}  are generally mixed and scattered, but with a mild thick disc overabundance comparing to thin discs. 
The overall picture of the sample collected by \citet{Bisterzo2006} from various sources is also in agreement with our observation. Thin disc stars (\citealt{Bisterzo2006} sample) are obviously [Cu/Fe] poorer than thick disc stars at the same metallicities; thick disc stars show a distinguishable trend and mild [Cu/Fe] increase up to [Fe/H]= -0.4~dex. 

For the metal-poor stars, \citet{Reddy2006, Reddy2008, Nissen2011, Ishigaki2013} observed scattered [Cu/Fe] abundances with a positive slope from [Cu/Fe]$\approx$-0.5~dex up to solar [Cu/Fe] at [Fe/H]$\approx$-1.0~dex. This agrees very
well with our results.  Furthermore, \citet{Nissen2011} showed that low-$\alpha$ metal-poor stars have significantly lower [Cu/Fe] than high-$\alpha$ stars. This was later confirmed by \citet{Ishigaki2013}. Cu from \citet{Nissen2011} was also reanalysed in LTE and NLTE by \citet{Yan2016} who showed that even if there are some differences in the LTE/NLTE abundances, the general behaviour is not changed. Our study strengthens these results.

Finally, [Cu/Fe] abundance ratios in EMP stars by \citet{Lai2008} (28 stars) show a tendency to decrease with decreasing metallicity. Our results for the most metal-poor stars show the same tendency.

We also point out that the [Cu/Fe] abundances in the bulge from \citet{Johnson2011} (156 giant stars) generally overlap that of the thin and thick discs. However these abundances scatter a lot (by $\approx$0.4~dex). Moreover, these abundance trends extend up to [Cu/Fe]$\approx$0.8~dex at [Fe/H]$\approx$0.5~dex.

Finally, the trend of [Cu/Mg] with respect to Mg (see Fig.~\ref{fig:EL-trends-mg}) is actually similar to those of [Fe/Mg] or [Ni/Mg]. As for these iron-peak species, the division between thin and thick discs are weak but real. The [Cu/Mg] ratio is found to stay constant and null for almost all thin discs, suggesting that the enrichment by supernovae of type-I did not produce any large amounts of copper. 

In Fig.~\ref{fig:EL-trends-cu}, we adopted copper as a reference element to study the [Mn,Fe,Ni, and Zn/Ni] trends. This figure shows that the evolution of copper, nickel. and ion are actually very similar in the [Cu/H]$>-1.0$~dex regime. All these three sequences indeed overlap and are close to [El/Cu]$\approx$0.0~dex. Moreover, whereas [Mn/Cu] and [Zn/Cu] show the exact opposite behaviour, the [Mn/Cu] thick disc trend is noticeably lower than the thin disc trend and the [Zn/Cu] thick disc trend is higher than in the thin disc. Finally, the picture in the metal-poor regime is rather different. Every [Mn,Fe,Ni, and Zn/Ni] high-$\alpha$ and low-$\alpha$ sequences behave similarly, i.e. [El/Cu] increases with decreasing of [Cu/H].

\subsection{Zinc}
 \label{sec:Zinc}
 
Zinc, as copper, is an intermediate element between the iron-peak and the neutron capture elements. It is produced by the same nucleosynthesis channels but with different contribution ratios.  
Up to half of the zinc is indeed spread out by type-II supernovae or hypernovae via  
$^{64}Zn$, which is produced from $\alpha$-rich freeze-out 
in $\gamma$ winds. The remaining part ($^{66,67,68,70}Zn$) is produced by the \textit{sr} process in massive stars (\citealt{Bisterzo2004}). Type-Ia supernovae and \textit{s} process occurring in AGB stars produce only a marginal part of zinc nuclei.
 
Our study reveals that the [Zn/Fe] versus [Fe/H] distribution of thick and thin disc stars show specific and distinct behaviours (see Figures~\ref{fig:ELEMENTS_separation} and \ref{fig:EL-trends}). The entire thick disc is obviously [Zn/Fe] richer than the thin disc. Moreover, the metal-poor part of the thick disc distribution is (red points) is tight and clearly separates from the thin disc at the same metallicities by about
0.2~dex whereas the thin disc [Zn/Fe] distribution is constant around the solar value, regardless of the  metallicity. 
These thin and thick disc distributions of [Zn/Fe] are very similar to the [$\alpha$/Fe] distributions (see top panel of Fig~\ref{fig:ELEMENTS_separation}). 
We therefore propose that zinc could be a very good species to chemically disentangle the thin and thick disc components
in the Galaxy.

The main $\alpha$-like behaviour of zinc in the discs was already marginally mentioned in some previous studies, which analysed uniform, but smaller samples (see, for instance the results of 172 FGK stars by \citealt{Mishenina2011} or 714 stars by \citealt{Bensby2014}) and also non-uniform data collections as in \citet{Bisterzo2006} (380 stars) and \citealt{Saito2009} (434 stars). \citet{Barbuy2015} showed that [Zn/Fe] in bulge stars also confirms its $\alpha$-like behaviour. However, the very clear separation between the thin and thick discs was never demonstrated before. 

On the other hand, the trend of [Zn/Fe] for the metal-poor $\alpha$-rich stars is close to the solar value up to [Fe/H]$\approx$-1.3 and then becomes slightly positive up to [Fe/H]$\approx$-1.0~dex, where it reaches supersolar [Zn/Fe] values and smoothly connects to the thick disc. The low-$\alpha$ sequence is generally lower than the high-$\alpha$
sequence (by $\approx$0.15~dex).
The largest uniform sample that can be used for comparison is that of 
\citet{Ishigaki2013} (97 stars), which showed very similar results that are 
also seen in the non-uniform samples of \citet{Bisterzo2004} and \citet{Saito2009}.
Finally, \citet{Nissen2011} also found that low-$\alpha$ metal-poor stars have significantly lower [Zn/Fe] ratios than high-$\alpha$ stars in their analyses. This tendency was confirmed by \citet{Ishigaki2013} and, again, agrees with our results based on a much larger and confident sample. According to \citet{Nissen2011}, NLTE corrections would lead to systematic changes in the derived [Zn/Fe] values by less than than $\pm$0.1~dex.
This conclusion is based on the work of \citet{Takeda2005} who estimated
the NLTE corrections for the three zinc lines selected for our analysis. Indeed, most of the NLTE-LTE abundance differences do not exceed 0.1~dex for the 4,722 and 4,810~\AA~lines. These differences
are mostly used to derive abundances for metal-poor stars, as shown by \citet{Takeda2005}, whereas the 6,362~\AA~line, mostly used in 
the [Fe/H]$>$-1.0~dex regime, is even less affected by such differences. According to \citet{Takeda2005,Takeda2016}, we thus do not expect Zn NLTE-LTE corrections to be larger than 0.03~dex and 0.06~dex for dwarfs and giants, respectively.

The [Zn/Fe] abundances of EMP stars by \citet{Cayrel2004} (70 stars) and \citet{Lai2008} (28 stars) show a clear tendency to increase with decreasing metallicity from [Zn/Fe]$\approx$0.15~dex at [Fe/H]$\approx$-2.5~dex to [Zn/Fe]$\approx$0.50~dex at [Fe/H]$\approx$-3.5~dex, although with a scatter as large as 0.25~dex. Our [Zn/Fe] results are also around -0.15~dex for metal-poor stars, which agrees with the literature.

The [Zn/Fe] abundances in bulge stars from \citet{Barbuy2015} (56 bulge giant stars) strongly decrease with increasing metallicity from 0.3~to~-0.5~dex in the [Fe/H] range from~-1.5~to~0.2~dex.
This [Zn/Fe] trend generally does not match any of our thin and thick disc trends.

Finally, the trend of [Zn/Mg] with respect to [Mg/H] (Fig.~\ref{fig:EL-trends-mg}, bottom panel) is very interesting. Both thin and thick disc trends overlap and they are close to [Zn/Mg]=0.0 for any metallicities. This actually confirms the $\alpha$-like behaviour of zinc in disc stars. In contrast, the metal-poor stars are found to be underabundant in [Zn/Mg], indicating that the production of zinc in massive metal-poor supernovae was smaller than the magnesium production at variance with the [Zn/Fe] of extremely metal-poor stars (EMPS) which is positive (e.g. \citealt{Cayrel2004}). 

We also refer to Appendix~\ref{sec:appendix1} for the [Mn,Fe,Ni, and Cu/Zn] ratios with respect to [Zn/H]. 
In this [El/Zn] versus [Zn/H] plane, the thick discs trends are lower and noticeably distinct than those of the thin discs. The strongest difference ($\approx$0.25~dex) is found for [Mn/Zn]. However, we point out that the high-$\alpha$ metal-rich and thin disc sequences
are much less distinctive in the [Mn,Fe,Ni, and Cu/Zn] ratios. Moreover, 
the metal-poor high-$\alpha$ trends are generally lower by $\approx$0.1~dex than those of the metal-poor low-$\alpha$ sequences for [Mn,Fe, and Ni/Zn] 
but they completely overlap for [Cu/Zn].
The adoption of zinc as a reference element is very rare in the literature, but it could be very promising, especially for the thin and thick disc separation. The last attempt to use zinc as a reference element was performed by \citet{Bihain2004} (38 FGK stars) who adopted the [Cu/Zn] ratio. Their results within error bars comply with our trends.

\section{Comparison with Galactic chemical evolution models}
\label{sec:models_comp}

The goal of this section is to compare the chemical abundance trends in the thin and thick discs
and in the halo presented above with some recent Galactic chemical evolution (GCE) models. 
Up to now, rather few GCE models have presented various abundance ratios of key elements in the different Galactic stellar populations in order to trace back their chemical history and better constrain all possible nucleosynthesis scenarios. 
Moreover, GCE models differ a lot depending on their internal formalism, assumptions, and input parameters (see \citet{Nomoto2013} for details).
For instance, one can encounter simple one-zone models, where it is assumed that the ISM has a uniform chemical composition, more realistic stochastic (e.g. Cescutti 2008) or hierarchical (Tumlinson 2006) models, and even more sophisticated 3D hydrodynamical simulations (e.g. \citealt{Kobayashi2011}). A variety of different inputs can also be found in such models as, for instance, the fraction of various types of supernovae and the stellar yields that are mostly sensitive to the various nucleosynthesis scenarios and timescales.

For our purpose, we selected the GCE models developed by \citet{Romano2010} and \citet{Kobayashi2011}
since they explicitly provide the Galactic evolution of a large set of species (up to Zn) separately for the different Galactic components (thin and thick discs and halo), which can be easily compared with our observed trends.

\subsection{Comparison with the \citet{Romano2010} models}
\label{sec:comparison_romano}
In \citet{Romano2010}, the two-infall model of \citet{Chiappini1997, Chiappini2001} was adopted. According to this assumption, the Galaxy was formed out of two main infall episodes. The inner halo plus thick disc component is formed during the first infall  on a relatively short timescale, 
whereas the thin disc is formed during the second infall. One can see the transition between these 
two phases in plots representing the evolution of a given chemical species ([X/Fe]) 
versus the iron abundance around [Fe/H]$\simeq$-0.6 (see figures 4-16 in \citealt{Romano2010}). 
Indeed, the merged halo plus thick disc sequence is seen at low metallicity up to [Fe/H]$<$-0.6 and is 
fully separated from the thin disc for [Fe/H]$>$-0.6.
Moreover, in their study, 
\citet{Romano2010} analysed the trends of 15 different chemical elements with known nucleosynthesis channels by testing different combinations of adopted stellar yields from low and intermediate mass stars (LIMS) and massive stars (see their Table~2 for details). The type-I supernovae yields by \citet{Iwamoto1999}, initial mass function (IMF) by \citet{Kroupa1993}, and the star formation rate (SFR) by \citet{Romano2005} were fixed for all these 15 models.

  \begin{figure}[htb]
   \advance\leftskip 0cm
   \includegraphics[scale=0.20]{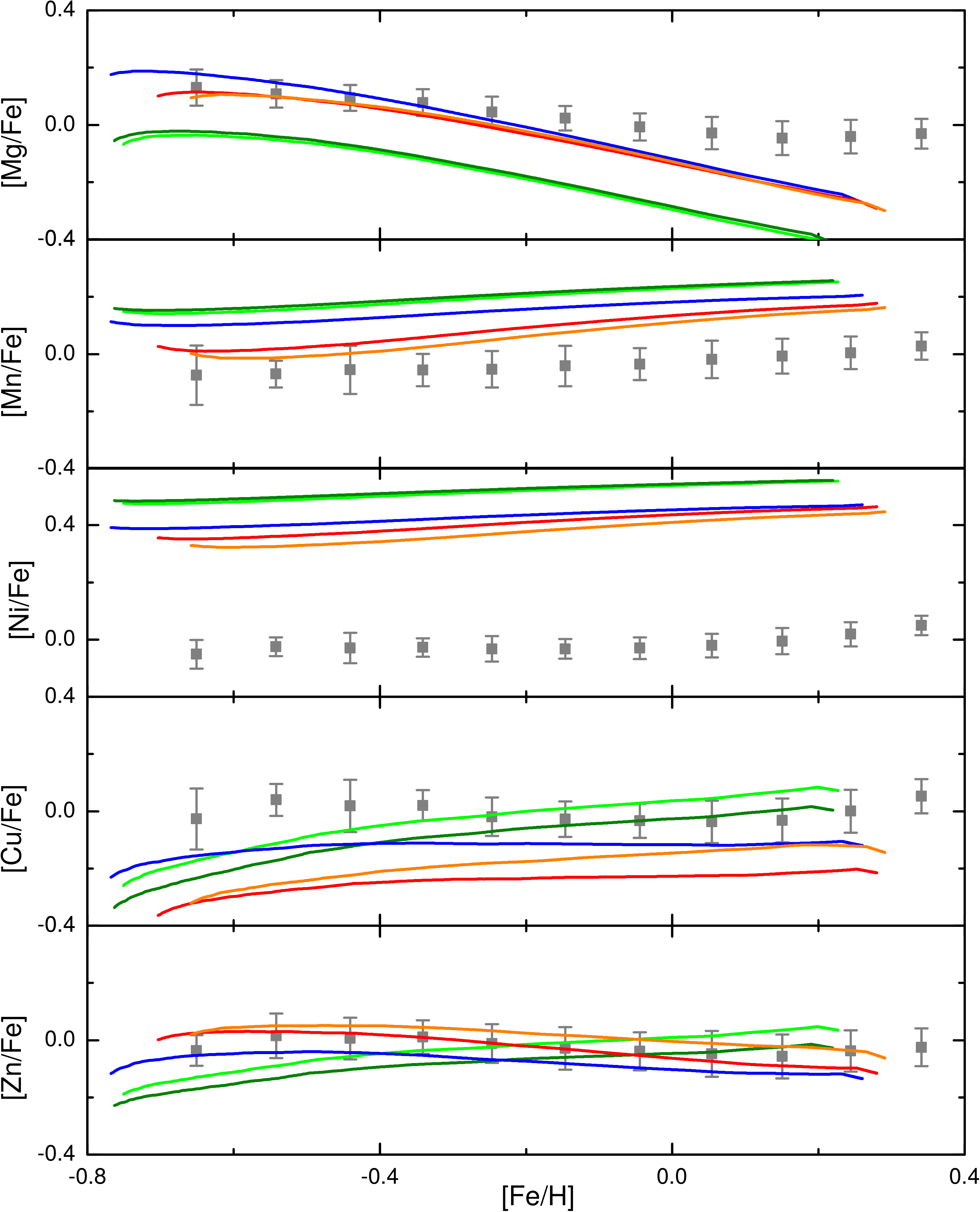}
  \caption{
 [X/Fe] as a function of [Fe/H] for the thin disc. Theoretical predictions are from the models of \citet{Romano2010}. 
 Models with adopted yields from \cite{Woosley1995} case A and \cite{Hoek1997} are shown as a light green line, the \cite{Woosley1995} case B and \cite{Hoek1997} are indicated with a dark green line, \cite{Kobayashi2006} ($\varepsilon_{\rm{HN}}=0$) and \cite{Hoek1997} are represented with a blue line, \cite{Kobayashi2006} ($\varepsilon_{\rm{HN}}=1$) and \cite{Hoek1997} are indicated with a red line, \cite{Kobayashi2006} ($\varepsilon_{\rm{HN}}=1$) plus Geneva pre-supernovae yields and \cite{Karakas2010} (orange line).
 The thin disc data points with error bars are the same as in Fig.~\ref{fig:EL-trends}.  }
  \label{fig:ROMANO_THIN_models}
  \end{figure}

  \begin{figure}[htb]
   \advance\leftskip 0cm
   \includegraphics[scale=0.20]{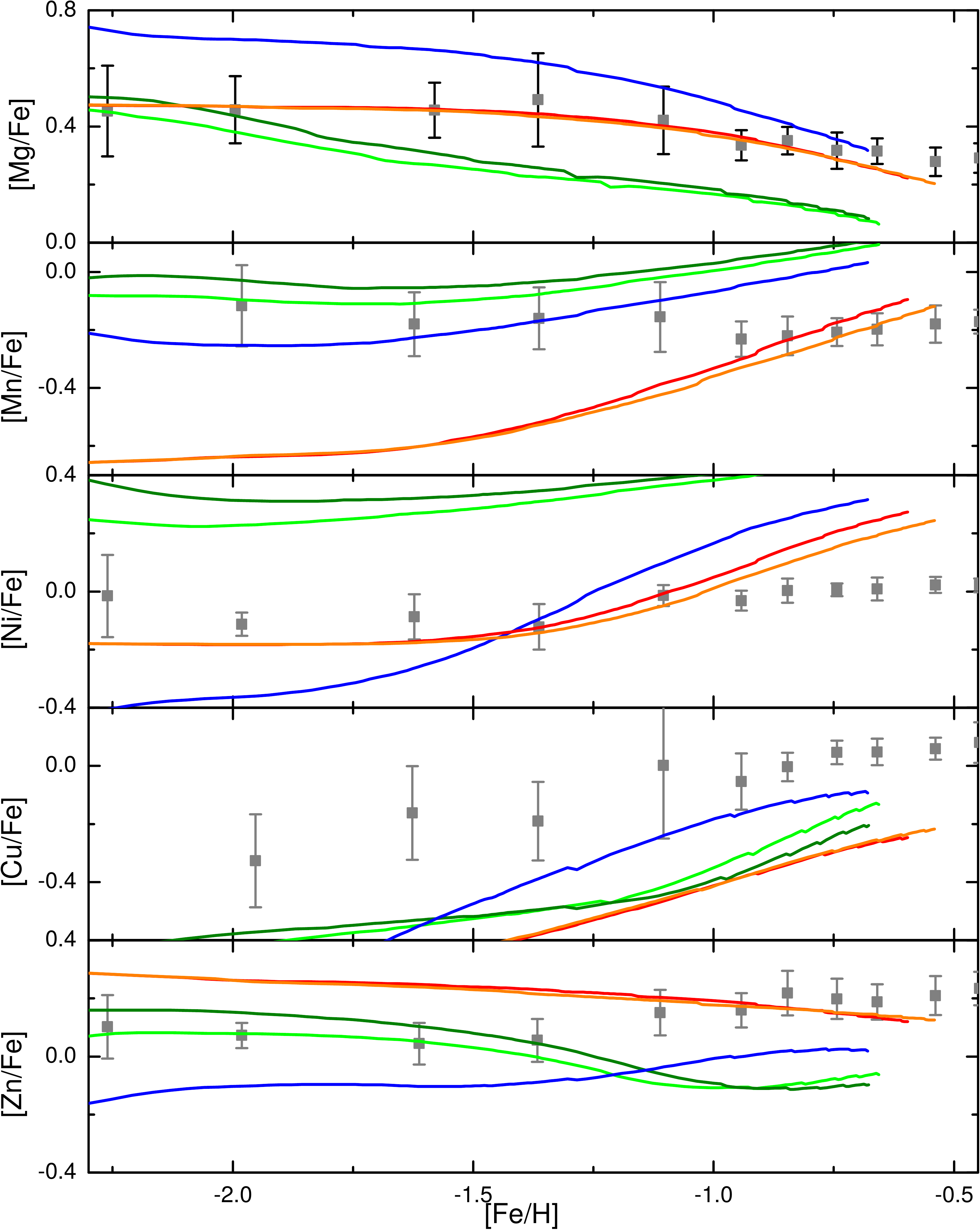}
  \caption{  Same as Fig.~\ref{fig:ROMANO_THIN_models}, but for the metal-poor sequence ([M/H]$<$-1.0~dex) and the thick disc.}
  \label{fig:ROMANO_HALOTHICK_models}
  \end{figure}

We compare the model results from these investigations with our observations in Figs~\ref{fig:ROMANO_THIN_models} and \ref{fig:ROMANO_HALOTHICK_models}. First, we point out that in their different models some trends deduced from different adopted yields are very similar to each other. For instance, the trends from their models 1 and 3, although adopting different yield assumptions, are almost indistinguishable.
Morever, the trends of the 5th to 15th models are very similar and vary  between two extreme cases (5th and 15th models). These 11 models differ in some adjustments of the massive star yields and different sources of LIMS yields. As a consequence, we decided to plot only models that actually show meaningful differences between each other in our Figs~\ref{fig:ROMANO_THIN_models} and \ref{fig:ROMANO_HALOTHICK_models}.
Generally, the choice of yields for the \citet{Romano2010} test was carried out in the way to show the most valuable modelled yield sets. The full listing of paired yields for LIMS and massive stars is shown in the caption of Figs~\ref{fig:ROMANO_THIN_models}. Here we briefly point out the most important physical differences between computations of the selected yields.

Massive stars are the main Galactic polluters, and thus the correct choice of their modelled yields are very important. For massive stars \citet{Romano2010} consider both cases (A and B) of the yields by \citet{Woosley1995}. They are computed without mass loss and without rotation. The case B of \citet{Woosley1995} corresponds to a slightly higher final kinetic energy of the ejecta at infinity than that of case A (typically 1.9~10$^{51}$ erg rather than 1.2~10$^{51}$ erg).
Another choice for massive stars is a modelled yield set by \citet{Kobayashi2006}, which were computed including metallicity-dependent mass loss and assuming that the fraction of stars that end their life as hypernovae is $\varepsilon_{\rm{HN}}=0$ and $\varepsilon_{\rm{HN}}=1,$ respectively. Some \citet{Kobayashi2006} yields are supplemented by pre-supernovae yields computed by the Geneva group\footnote{Yield sources of the Geneva group: \citet{Meynet2002, Hirschi2005, Hirschi2007, Ekstrom2008}.} with both mass loss and rotation.

 During thermally pulsing AGB (TP-AGB) phase, LIMS experience a very rich nucleosynthesis whose
products are convected to the outermost layers and eventually injected into the ISM by stellar winds. The correct choice of LIMS yields is crucial for modelling of the evolution of some elements. Most of the models that we show in in Figs~\ref{fig:ROMANO_THIN_models} and \ref{fig:ROMANO_HALOTHICK_models} employed yields of \citet{Hoek1997}, which are constructed with metallicity-dependent tracks, including the effects of the first and second dredge-ups and hot bottom burning (HBB) on the yields. Extra-mixing was not considered by \citet{Hoek1997}. Another choice of LIMS yields was the most recent study by \citet{Karakas2010}, who evolved LIMS models from the zero-age main sequence to near the tip of the TP-AGB, included dredge-ups and HBB. The contribution from  extra pulses was not included.

%
%
%
%
%

The comparison between these modelled trends and our thin disc data (Fig.~\ref{fig:ROMANO_THIN_models})
leads to the following conclusions. It can be seen that 
the magnesium trends for the thin disc adopting the yields from \citet{Woosley1995} underestimate Mg abundances of the thin disc stars by about 0.2~dex. 
On the contrary, the \citet{Kobayashi2006} yields reflect these abundances [Mg/Fe] much better, 
although they fail to reproduce the metal-rich ([Fe/H] $>$ -0.2) part of the thin disc.
The predicted slope of [Mg/Fe] versus [Fe/H] is indeed too large, leading to a discrepancy of about
0.2~dex at [Fe/H] $\sim$ +0.2.
On the other hand, the predicted manganese and nickel in the thin disc are overestimated by all models because
the disagreement is very large for nickel, reaching more than 0.4~dex. 
However, it should be pointed out that the models assuming a hypernovae fraction $\varepsilon_{\rm{HN}}=1$ are closer to the observations.
On the contrary, copper is underestimated by all models based on the \citet{Kobayashi2006} yields in the thin disc. However, the agreement with the \citet{Woosley1995} yields (green lines) is better, although
they cannot reproduce the slope of the thin disc data trend.
Finally, zinc is nicely reproduced by the models based on the \citet{Kobayashi2006} yields, particularly
when the hypernovae yields are included, whereas 
the \citet{Woosley1995} yields fail to reproduce the global Zn shape, where the largest disagreement is found for metal-poor stars.

As for the halo plus thick disc comparison (see Fig.~\ref{fig:ROMANO_HALOTHICK_models}),
it is first important to note that the model results strongly differ from the hypernovae fraction 
$\varepsilon_{\rm{HN}}$ of the \citet{Kobayashi2006} models (blue and red/orange trends in Fig.~\ref{fig:ROMANO_HALOTHICK_models}). Actually, this is not surprising since massive stars with short lifetimes are expected to be 
the main (and almost sole) sources of chemical elements in the old and metal-poor populations. 
It can be seen that the magnesium behaviour with iron is rather well reproduced by models based on the \citet{Kobayashi2006} yields and a hypernovae fraction $\varepsilon_{\rm{HN}}=1$ (i.e. all stars that explode are hypernovae). 
On the contrary, the [Mn/Fe] abundance ratio trend is not perfectly fitted by any model.
The models adopting the \citet{Kobayashi2006} yields and $\varepsilon_{\rm{HN}}=1$ could reproduce the thick disc data, but strongly underestimate Mn abundances for [Fe/H]$<-1$. Adopting $\varepsilon_{\rm{HN}}$ = 0 leads to an overestimation of [Mn/Fe], but the model is closer to the observations for
metal-poor stars ([Fe/H]$<-1$). In contrast, the \citet{Woosley1995} yields lead to closer
agreement for the metal-poor regime ([Fe/H]$<-1$) and an overestimation for the metal-rich regime.
Nickel abundances could be partially reproduced when [Fe/H]$<-1$ only by models based on the \citet{Kobayashi2006} yields regardless of the value of $\varepsilon_{\rm{HN}}$. It seems that an adjustment of $\varepsilon_{\rm{HN}}$ brings the models based on the \citet{Kobayashi2006} yields slightly closer to the data for the thick disc (more metal-rich) regime although the agreement is poor. We also point out
that the models based on the \citet{Woosley1995} yields drastically overestimate the nickel abundances
at any metallicity.
Furthermore, the copper abundances are also very badly reproduced (strongly underestimated) by every model,
regardless of which yields, IMF and/or hypernovae rate assumptions are considered. 
Finally, we point out that zinc abundances cannot be fully reproduced by any model either. The \citet{Kobayashi2006} yields and $\varepsilon_{\rm{HN}}=1$ work well enough for the thick disc, but overestimate [Zn/Fe] by at least 0.2~dex at the [Fe/H]$<-1$ metal-poor regime.
Furthermore, one could see that varying $\varepsilon_{\rm{HN}}$ between 0 and 1 leads to
strong differences, revealing the importance of the actual sources of the yields to distengangle
the chemical evolution of these Galactic populations.
On  the other hand, the \citet{Woosley1995} yields predict the zinc trend for 
[Fe/H]$<-1.3$ regime rather well, but also predict an increase at lower metallicity that
is not at all observed.

Summarising these comparisons between our observed chemical trends and the \citet{Romano2010}
predicted trends in the different Galactic components, 
one can conclude that no single combination of the adopted stellar yields is able to 
reproduce well all our observed trends of the chemical abundances ratios with respect to metallicity. 
We could however favour the model adopting the massive star yields of \citet{Kobayashi2006} with a large ratio of hypernovae together with the yields of \citet{Karakas2010} for the LIMS. These models can indeed partially reproduce well most of the magnesium trends in the metal-poor sequence and the discs 
(except for metal-rich thin disc stars). Nevertheless, we point out below the most problematic areas in this comparison.

The typical [$\alpha$/Fe] versus [Fe/H] trend is present in our observations of 
the magnesium abundances. One could see the classical flattening in the [Mg/Fe] ratio
for [Fe/H]$>-0.2$, indicating that the contribution of type-II and type-I supernovae are in equilibrium beyond this point. This typical [$\alpha$/Fe] versus [Fe/H] trend is a fingerprint of isolated stellar system, providing insight into their chemical enrichment history. Since the shape and position of the trend is controlled by the SFR and IMF,
we point out that none of the present models can reproduce the flat part of the metal-rich thin disc regime [F/H]$>-0.2$. 
We also note that all the 15 \citet{Romano2010} models are based on the same IMF (\citealt{Kroupa1993}) and SFR (\citealt{Romano2005}). This means that any differences 
between these 15 models can only result from the different adopted yield sources.
However, our favoured model complies with the magnesium trends very well from [Fe/H]$=$-2.5 up to [Fe/H]$=$-0.2 although it does not reproduce the flattening above [Fe/H]$=$-0.2. Therefore, there could be a source of some insights on the thin-disc evolution. 
Some possibilities of such a disagreement can be suggested. 
One could mention the different roles of the IMF and SFR in the thin disc. 
It might be that some models do not adopt realistic estimate numbers of type-II/type-Ia supernovae that cause the behaviour of $\alpha$ versus iron trends. This issue could also be solved by considering a significantly larger contribution from the LIMS. The yields for magnesium are indeed smaller by a factor 4-5  in LIMS than in type-II supernovae (e.g. \citealt{Karakas2010} and \citealt{Kobayashi2006}). Thus it is unsafe to address this issue to the magnesium production from LIMS. Most likely, the combination of all factors could flatten the [Mg/Fe] trend for metallicities larger than -0.2~dex.

The behaviour of the zinc abundances is also predicted rather well by our
favoured model, except in the case of the thick disc. 
The zinc chemical evolution models based on \citet{Kobayashi2006} and \citet{Karakas2010} yields show satisfactory results for both discs, but not for the metal-poor regime. The zinc production mechanisms in very massive first generation stars or metal-poor type-II supernovae should be investigated more deeply.
Our favoured model is also the closest to the manganese trends 
in the thin disc, although there is an agreement only for metal-rich high-$\alpha$ sequence stars. 
As a consequence, all models overestimate the manganese content of the thin disc. 
The solution may lie either in a contribution by type-I supernovae that is  too large, or in an excessive production of Mn from type-I supernovae at late epochs.
This was demonstrated by \citet{Cescutti2008b}, who studied
Mn evolution in the solar neighbourhood, the Galactic bulge, and
the Sagittarius dwarf spheroidal galaxy. Considering this, metallicity-dependent yields of Mn from type-Ia supernovae could provide a solution to
the problem of the overproduction of Mn at late epochs.

On another hand, it seems that the behaviour of nickel is typically difficult to model. The problematic jump of modelled trends for [Fe/H]$>-1.0$ regime is due to the type-I supernovae input. This could be solved by varying the number of electrons per nucleon (electron mole number) $Y_{\rm{e}}$ (see \citealt{Nomoto2013}). However it may be difficult to solve this problem because Ni and Zn have an opposite response to $Y_{\rm{e}}$ as was pointed out by \citet{Sneden2016}. Another way to start solving the nickel issue could be the tuning of the propagation
speed of the burning front and the central density of the white dwarf (\citealt{Iwamoto1999}) or 3D simulations of thermonuclear explosions (\citealt{Ropke2012}).

Finally, the underproduction of copper in the models based on \citet{Kobayashi2006} and \citet{Karakas2010} yields should be investigated. The $s$ process acting in low-mass AGB stars is not included in their models although we could expect a significant contribution of copper by these stars. We already pointed out in Sec.~\ref{sec:Copper} that the input of $s$ process occurring in AGB stars should be small, but such an input must be further investigated before stating any conclusions. For example, the contribution by the $s$ process to the solar copper abundances should be around 27\% in total, but only 5\% is expected to come from AGB stars while the largest amount should come from massive stars by weak s process (\citealt{Bisterzo2004}).

\subsection{Comparison with the Kobayashi et al. models}
\label{sec:comparison_kobayashi}
These authors developed a one-zone GCE model for the star formation histories of the solar neighbourhood, halo, thick disc, and bulge.
The two most recent published versions of these models (\citealt{Kobayashi2006, Kobayashi2011}) are rather similar. 
For the halo, they use an outflow model without infall where the star formation efficiency is much lower than in the other Galactic substructures. For the thick disc, an infall plus wind model was adopted over a relatively short timescale (3~Gyr). In contrast, the star formation episode in the solar neighbourhood (i.e. for the thin disc) lasts for 13~Gyr. On another hand,
the combination of yields adopted by these authors includes a variety of enrichment sources: yields from stellar winds, AGB and super AGB stars, core collapse supernovae, rotating massive stars, pair-instability supernovae, and type-I supernovae. 
The differences between the two model versions consist in recomputed yields and updated initial mass functions (see \citealt{Kobayashi2000, Kobayashi2006, Kobayashi2011} for details). We compare their model results with our observations in Figs.~\ref{fig:thin_models}, ~\ref{fig:thick_models}, and \ref{fig:HALO_models} for the thin disc, thick disc, and halo components, respectively.

Three models are compared for the thin disc in Fig.~\ref{fig:thin_models}: 
the 2006 and 2011 models that do not include rotating massive stars (red and blue lines respectively) and the 2011 model with rotating massive stars (dashed blue line). As  \citet{Kobayashi2011} explain, for the elements heavier than sodium, the difference (which is very large for Zn and Cu) among the models with and without rotating massive stars is caused by the difference in the IMF. Indeed, these authors set a stellar mass 50$_\odot$ as the upper limit of core-collapse supernovae mass, but a limit at 120$_\odot$ is adopted in the case of rotating massive stars. 

Figure~\ref{fig:thin_models} shows that the magnesium trend is reproduced rather well
by the \citet{Kobayashi2006} model for [Fe/H]$>$-0.5, whereas the \citet{Kobayashi2011} models overestimated the [Mg/Fe] ratio by about 0.1~dex. All models underestimate the manganese abundances, although \citet{Kobayashi2006} could agree within 1$\sigma$. Nickel is always strongly overestimated by 0.1-0.3~dex at any metallicities metallicity. The copper abundances of \citet{Kobayashi2006} are partially in agreement for [Fe/H]$>$-0.3, but the trend at lower metallicity is not fully predicted. \citet{Kobayashi2011} model without rotating massive stars overestimated copper by about 1.0$\sigma$~--~1.5$\sigma$. Finally, the \citet{Kobayashi2006} model  well reproduces the zinc behaviour in the thin disc, contrary to \citet{Kobayashi2011} who overestimated it. Interestingly, the \citet{Kobayashi2011} models that include rotating massive stars reproduce copper and zinc abundances rather well in the thin disc.

 \begin{figure}[htb]
   \advance\leftskip 0cm
   \includegraphics[scale=0.30]{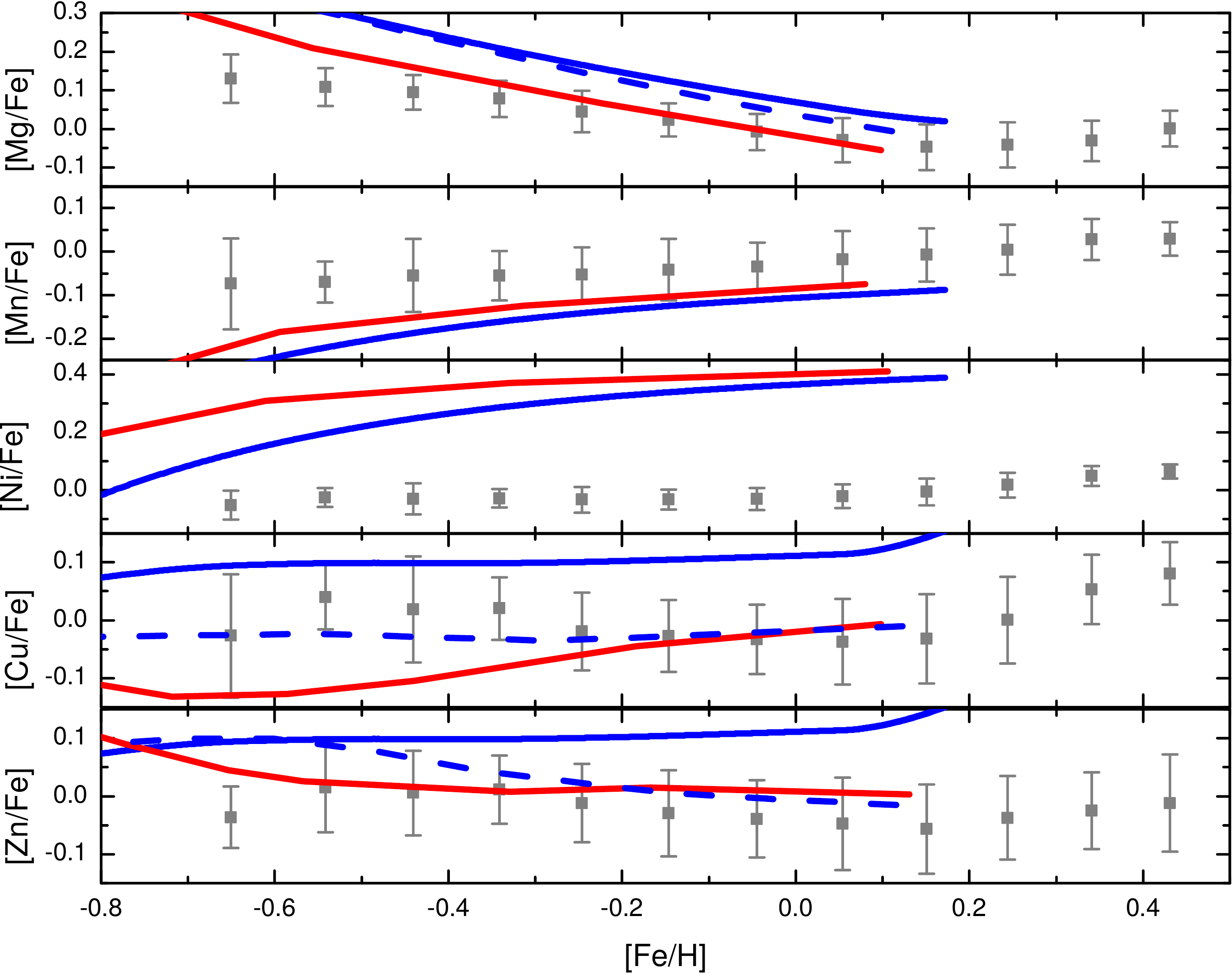}
  \caption{
 [X/Fe] as a function of [Fe/H] for the thin disc. Theoretical predictions from the models of \citet{Kobayashi2006} are shown as the red line, \citet{Kobayashi2011} is indicated with a blue line and \citet{Kobayashi2011} with massive star rotation is indicated with a dashed blue line.  The thin disc data points with error bars are the same as in Fig.~\ref{fig:EL-trends}.
}
  \label{fig:thin_models}
  \end{figure}

   \begin{figure}[htb]
   \advance\leftskip 0cm
   \includegraphics[scale=0.30]{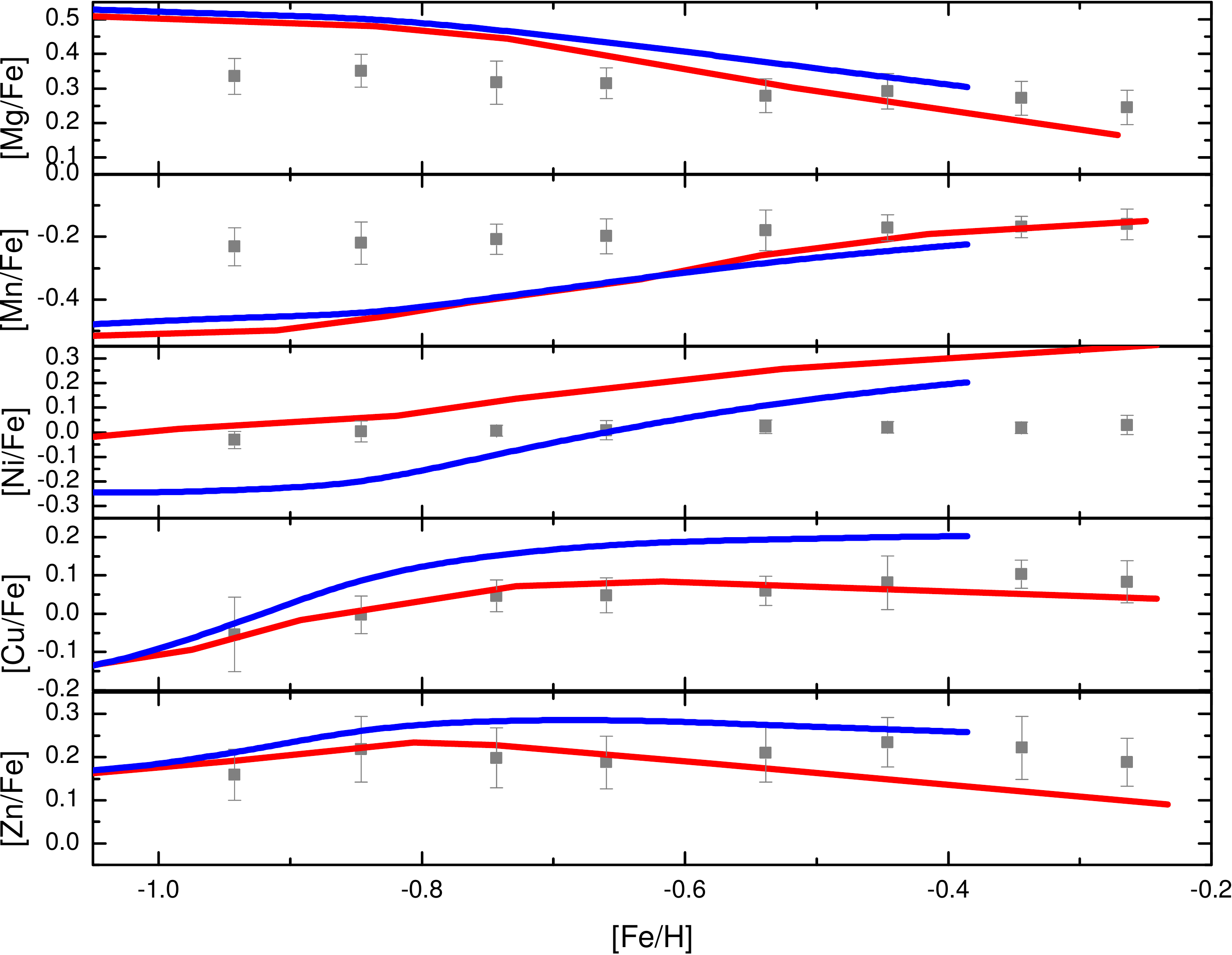}
  \caption{  Same as Fig.~\ref{fig:thin_models}, but for the thick disc.}
  \label{fig:thick_models}
  \end{figure}

  \begin{figure}[htb]
   \advance\leftskip 0cm
   \includegraphics[scale=0.30]{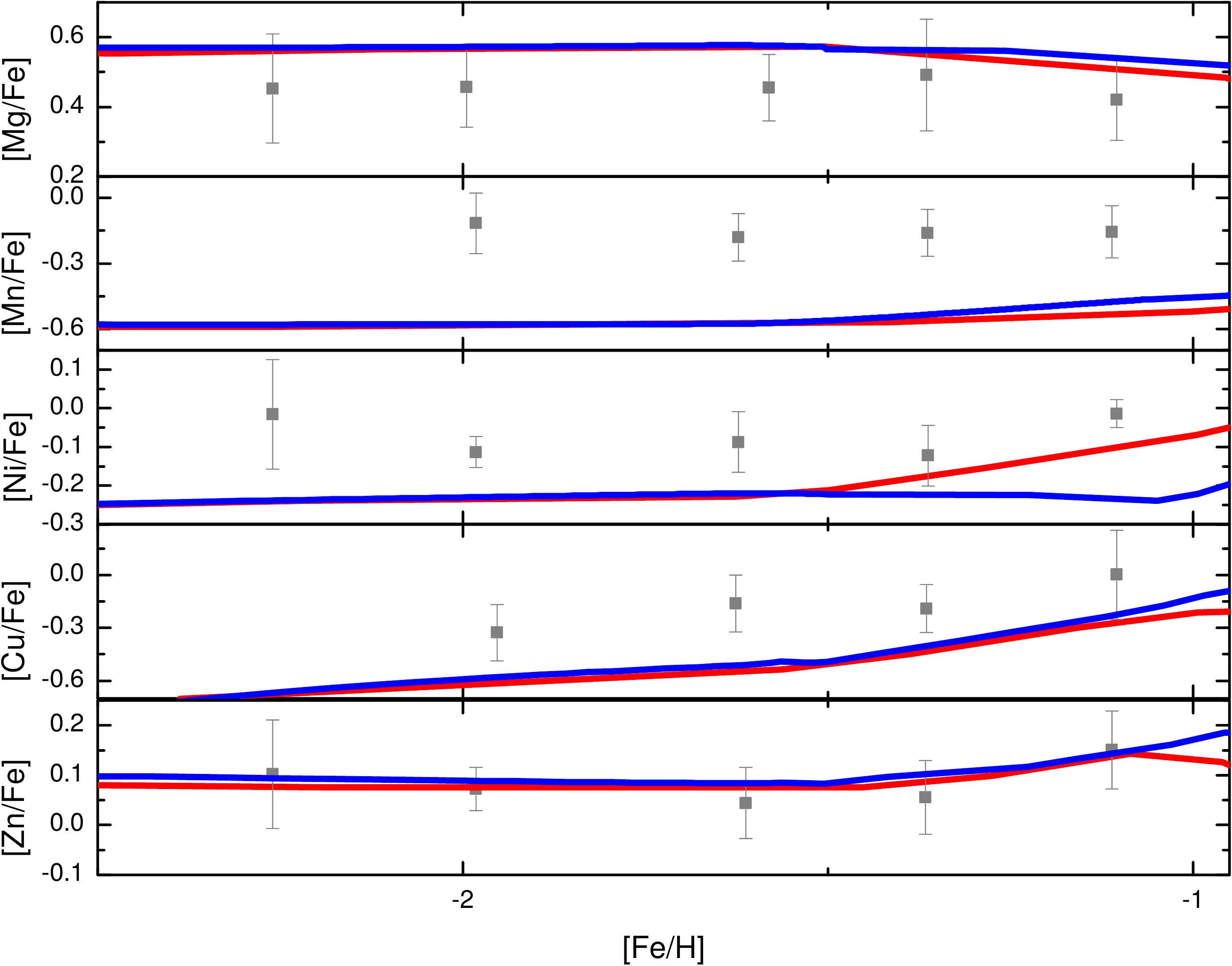}
  \caption{  Same as Fig.~\ref{fig:thin_models}, but for the metal-poor sequence.}
  \label{fig:HALO_models}
  \end{figure}

Regarding the thin disc and halo, only the 2006 and 2011 models without rotating massive stars are available (red and blue in Fig.~\ref{fig:thick_models}, respectively). The thick disc abundance trends (except nickel) are partially reproduced by both models (see Fig.~\ref{fig:thick_models}).   Furthermore, magnesium is overestimated and manganese underestimated in the metal-poor part
([Fe/H]$<$-0.6). Nickel is badly estimated by both models, except in the [Fe/H]$<$-0.8 regime
by \citet{Kobayashi2006}. \citet{Kobayashi2006} reproduced the copper trend very well, contrary to the prediction of \citet{Kobayashi2011}. Finally, zinc is partially reproduced by both models,
but a better fit could be obtained with an intermediate model between those of 
\citet{Kobayashi2006} and \citet{Kobayashi2011}. 

Finally, the metal-poor abundance trends (see Fig.~\ref{fig:HALO_models}) are reproduced rather well  
for magnesium and zinc by both models. In contrast, the manganese, nickel, and copper abundances
are strongly underestimated by 0.2~dex~--~0.3~dex.

In summary, the global agreement between these models and the observations are rather good, although several problems can be pointed out. First, it seems that the newer \citet{Kobayashi2011} model tends to show a stronger disagreement with the observations than the \citet{Kobayashi2006} older model. 
The most significant differences between models of both generations are in the high-metallicity regime at [Fe/H]$>-1.0$, when the type-I supernovae start to enrich ISM. We point out two reasons such a difference could be produced in the modelled trends: the modelled enrichment of ISM by type-I supernovae and/or the choice of the IMF.
First, both models used the type-I supernovae yields of \citet{Nomoto1997b}, however the treatment of type-I supernovae (e.g. lifetime distribution calculations) were updated. The different SFRs, MDFs, and 
also lifetime distribution calculations consequently create differences in chemical enrichment models beyond [Fe/H]$=-1.0$ (see \citealt{Kobayashi2000} and \citealt{Kobayashi2009} for a detailed description).
Secondly, models of both generations were created with different IMF. \citet{Kobayashi2006} adopted the \citet{Salpeter1955} IMF whereas \citet{Kobayashi2011} adopted the IMF of \citet{Kroupa2008}. The significance of the IMF was already show in Fig.~\ref{fig:thin_models}, where the difference mismatch between the blue and dashed blue lines are only caused by the upper limit of the core-collapse supernovae mass in the IMF.

Then, both models have difficulties reproducing the manganese and nickel
trends in all the Galactic populations. Nickel is probably the most problematic element to model since it is predicted to react as manganese, but this is clearly not observed. It seems that yields from core-collapse supernovae should be a bit higher both for manganese and nickel to meet observations. Both \citet{Kobayashi2006} and \citet{Kobayashi2011} modelled the jump of [Ni/Fe] at [Fe/H]$\approx$-1.0 followed by a large disagreement with observations (as in \citet{Romano2010}). This is probably caused by modelled overproduction of nickel by the type-Ia supernovae (see the discussion in~Sec.~\ref{sec:comparison_romano}).

\subsection{Comparison with the \citet{Kubryk2015} models}
\label{sec:kubryk_kobayashi}
\citet{Kubryk2015} created CGE models with several new or updated ingredients: atomic and molecular gas phases, star formation depending on the molecular gas content, recent homogeneous yields, and observationally inferred type-I supernovae rates. Their model is based on a updated version of the “independent-ring” model for the MW presented in \citet{Prantzos1995,Prantzos1996}, and \citet{Boissier1999}. These authors adopted the most recent stellar yields from \citet{Nomoto2013}, which actually consist in an updated version of the \citet{Kobayashi2006, Kobayashi2011} yields. However, their adopted yields are forced to match the solar composition for 4.5~Gyr-old stars located in the solar vicinity. Moreover, an interesting addition is the inclusion of radial migration with parametrised time- and radius-dependent diffusion coefficients based on the analysis of N-body simulations. These simulations
indeed include epicyclic motion (blurring) and radial migration (churning) following 
\citet{Sellwood2002} and \citet{Schonrich2009}. Finally, they also considered parametrised radial gas flows, induced by the action of the Galactic bar.


We compare the CGE model results of \citet{Kubryk2015} with our observations in Figs.~\ref{fig:KUBRYK_THIN_models} and \ref{fig:KUBRYK_THICK_models} for the thin and thick discs, respectively. 
Since these authors provide trend predictions for chemical elements up to nickel (from [Fe/H]$\approx$-1.0~dex to 0.05~dex for the thick disc and from [Fe/H]$\approx$-0.8~dex to 0.45~dex for the thin disc) the comparison is only available for magnesium, manganese and nickel because these two last species are hardly modelled (see previous subsections). \citet{Kubryk2015} adopted a threshold at 9~Gyr to define thin (younger than 9~Gyr) and thick (older than 9~Gyr) discs in their models.
One could see that magnesium is nicely modelled for both discs. Manganese and nickel are also modelled well for the thin disc, although
the slight increase of nickel with metallicity is not predicted fully when [Fe/H]$>$0.2~dex.
However, in the thick disc, manganese is overestimated by $\approx$0.1~dex and nickel is generally underestimated by $\approx$0.15~dex.

Moreover, another important observational constraint provided by our observations can be seen in the rather low chemical dispersions
reported in the thick disc. Our Fig.~\ref{fig:EL-scatter} shows that dispersions as small as $\approx$0.05~dex
are found for Mg, Mn, and Ni (particularly in the metal-poor part). 
\citet{Kubryk2015} actually simulated the dispersion of the thick disc trend for [$\alpha$/Fe] (see their Fig.~14) and  predict dispersions as high as 0.09-0.12~dex for [Fe/H]$<$-0.6~dex (as estimated directly from their figure). This is in disagreement with our observed low dispersions ($\approx$0.05~dex), which could be an indication of a rather inefficient stellar migration for the oldest stars in the thick disc.

In summary, it is obvious that \citet{Kubryk2015} CGE models generally work better for Mg, Mn, and Ni than those studied in the previous subsections.
Some probable reasons for this better agreement could be: 
(i) \citet{Kubryk2015} used newer yields from \citet{Nomoto2013},
(ii) the Kubryk et al. adopted yields (\citealt{Nomoto2013}, corrected yields) are forced to match the solar composition for 4.5~Gyr-old stars located in the solar vicinity, and
(iii) Kubryk et al. considered radial migration ("blurring" and "churning") in their models. 
The rather small
dispersions of the chemical abundances observed in our sample could be 
consistent with inefficient migration processes.

  \begin{figure}[htb]
   \advance\leftskip 0cm
   \includegraphics[scale=0.30]{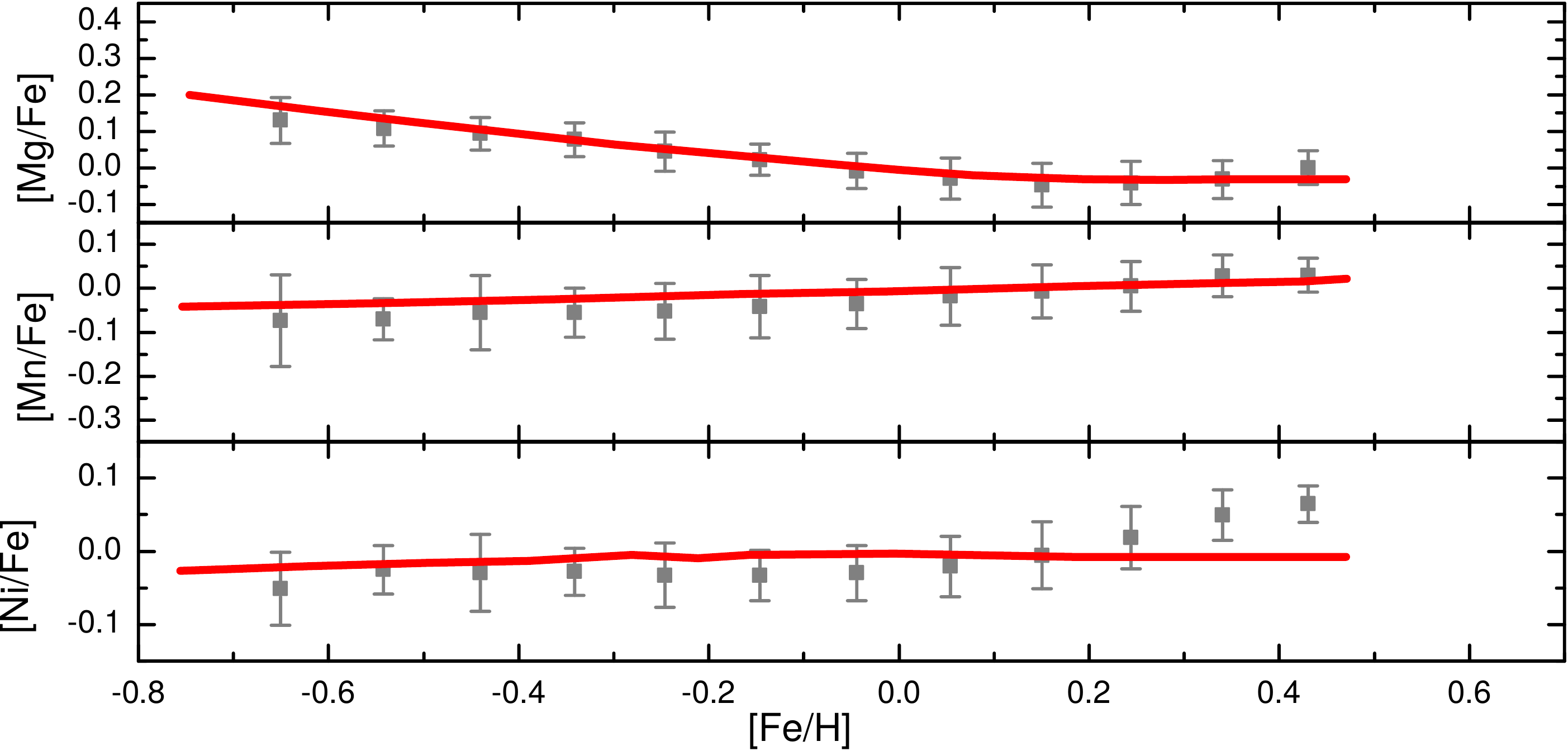}
  \caption{
  [X/Fe] as a function of [Fe/H] for the thin disc. Theoretical predictions from the models of \citet{Kubryk2015} are plotted as a red line. The thin disc data points with error bars are the same as in Fig.~\ref{fig:EL-trends}. }
  \label{fig:KUBRYK_THIN_models}
  \end{figure}

  \begin{figure}[htb]
   \advance\leftskip 0cm
   \includegraphics[scale=0.30]{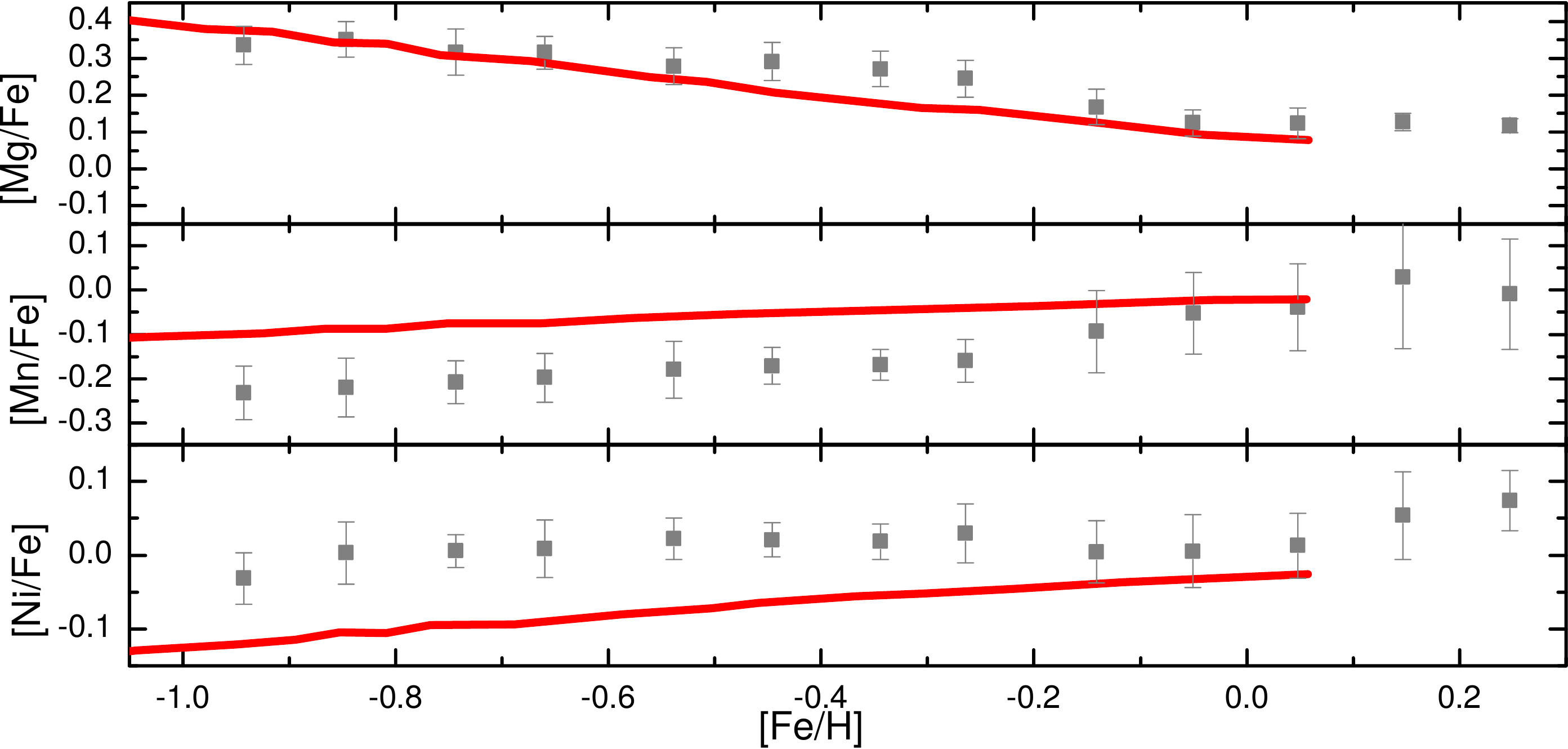}
  \caption{ Same as Fig.~\ref{fig:KUBRYK_THIN_models} but for the thick disc sequence.}
  \label{fig:KUBRYK_THICK_models}
  \end{figure}

\section{Summary}
\label{sec:conclusion}

Our study has extended previous efforts devoted to the iron-peak element chemical characterisation of Galactic populations. We have  homogeneously derived manganese, iron, nickel, copper, and zinc elemental abundances, together with magnesium as a reference element, for a large sample of 4\,666 FGK non-rotating giant and dwarf stars. As already revealed in previous studies (e.g. \citealt{Adibekyan2012,Recio2014}),
magnesium shows a bimodal distribution typical of other $\alpha$ elements but with a much better contrast.
The [Mg/Fe] ratio has thus been adopted to define chemically the three distinct populations (for [Fe/H]$>$-1.0~dex) that we attributed to the thin
disc, thick disc, and metal-rich high-$\alpha$ sequence. We then explored the general behaviour of Mn, Ni, Cu, and Zn abundances 
(in [El/Fe]~versus~[Fe/H] planes) in these main different components. We showed that
nickel and copper do not show any clear separation between both discs. However, it seems that the thick disc should be slightly richer in Ni and Cu than in the thin disc. Zinc is found to be very clearly separated between both discs with a behaviour that is very similar to magnesium. 
Moreover, looking at a [Zn/Mg]~versus~[Mg/H] plane, we showed that the trends of both the thin and thick disc overlap and that
they are close to [Zn/Mg]=0.0 for any metallicities. This result was expected since Zn should be an $\alpha$-like element from
a stellar yield point of view, but this was never clearly observed before in a large homogeneous sample. The [Zn/Fe] ratio can therefore be easily used to tag the different Galactic populations and Zn in the Galaxy is probably mostly produced by type-II supernovae. An opposite behaviour was revealed for manganese. The
thick disc is found to be slightly poorer in Mn (by $\approx$ 0.03-0.05~dex) than the thin disc. 
Moreover, it has been shown in a [Mn/Mg]~versus~[Mg/H] plane that the manganese-to-magnesium ratio could improve the efficiency of chemical tagging for metal-rich stars ([Fe/H]$>$-1.0~dex). And, since zinc behaves as an $\alpha$-like element, the [Mn/Zn] ratio could even be adopted if Mg abundances are not available.

Although our sample contains fewer metal-poor stars (144 stars for [Fe/H]$<$-1.0~dex), we confirm the distinction between two metal-poor populations previously revealed by \citet{Nissen2010} and based on their $\alpha$-element abundances. The distinction is also nicely seen in copper and zinc abundances, but it is not statistically visible in manganese and nickel. According to these authors, the metal-poor, $\alpha$-poor population stars could be former members of accreted satellite galaxies, where yields from various supernovae types were different than in our Galaxy, as is seen in several studies of the smaller satellite galaxies \citep[see for example][and references therein]{Tolstoy2009,Swaelmen2013}. 
If we follow this idea of accreted satellite galaxies, one could try to constrain their chemical properties. [Mg/Fe], [Cu/Fe], and [Zn/Fe] (mostly produced by massive stars) ratios are generally lower than those of the Milky Way disc stars. This could suggest a smaller contribution of massive stars or a slower enrichment process during the chemical evolution of these accreted galaxies.

In order to better understand the chemical evolution of the Galaxy, we compared our observational
results to recent realistic models that predict the chemical trends of the Galactic main components. In the literature, one can find many GCE models that differ a lot depending on their internal formalism, assumptions, and input parameters (e.g.~\citealt{Cescutti2008,Tumlinson2006,Nomoto2013}). However only some recent models (\citealt{Romano2010,Kobayashi2006,Kobayashi2011,Kubryk2015}) provide theoretical evolutionary trends separately for the thin and thick discs and/or the halo and for a variety of chemical elements (up to Z=30). 
The comparison between the observed and theoretical trends led to the conclusion that it is still difficult for models to reproduce
the manganese and nickel behaviours in most of metallicity domains. However, even if adjustments in some metallicity domains are still required, magnesium, zinc, and partially copper are generally modelled rather well. Finally, models adopting the \citet{Kobayashi2006} yields for type-II supernovae generally show better agreement with observations compared to the other models discussed in Sections~\ref{sec:comparison_romano} and \ref{sec:comparison_kobayashi} of this paper. The stellar yields, which still suffer from large uncertainties should be probably updated. This was clearly demonstrated by \citet{Kubryk2015}  who adopted normalised yields according to solar chemical properties, leading to a significantly better agreement with our observational data. 
This test actually stresses the importance of such large high-quality observational studies to create realistic models of Galactic evolution.

On another hand, the comparison of chemical evolution history in the solar neighbourhood and other populations (e.g. bulge or dSphs) could provide a better understanding about the chemical enrichment sources of our elements of interest. Among our three sources of models, only \citealt{Kobayashi2006,Kobayashi2011} provide the specific bulge model and if we compare to the Mn and Zn results of \citet{Barbuy2013,Barbuy2015}, we see that models are in strong disagreement with observations. This stresses the need to fit models in independent stellar systems (e.g. both discs, halo, bulge, and dSphs) simultaneously using the same nuclear prescriptions.
A good example is the model by \citet{Cescutti2008b}, who were able to fit the solar neighbourhood, bulge, and Sagittarius dSph galaxy systems.  Consequently, these authors emphasised quantitatively, by means of Galactic chemical evolution models, that the type-Ia supernovae Mn yields must depend on metallicity. This was later supported by models of \citet{Romano2011}. 


Finally, we note the fact that the present observational trends in the studied iron-peak elements are very tight ($\approx$0.05~dex). This provides another observational constrain for Galactic evolutionary models that includes some kind of radial redistribution of stars over time. The efficiency of the migration should indeed satisfy the observational dispersions of stellar populations. Our observations could favour a rather low efficiency for stellar migration.

In summary, our large uniform spectroscopic data set allows us to learn a lot about the Galactic populations in the solar vicinity. With such significant statistics we can indeed describe trends of chemical abundance ratios and make a direct comparison with models, providing strong modelling constraints. 
As a consequence, this work shows that that, in the Gaia era, detailed chemical analyses of large high-resolution datasets can provide unique information about Galactic population studies.

\begin{acknowledgements}

Part of the calculations
were performed with the high-performance computing facility MESOCENTRE, hosted by
OCA. We are grateful to Nicolas Prantzos for useful suggestions that improved the content
of this article. We thank the anonymous referee for her/his constructive comments.
T. Masseron and B. Plez were particularly kind in providing us with some molecular data.
We appreciate that D.~Romano, C.~I.~Johnson, and C.~Kobayashi kindly shared their model data.
This work was partly supported (\v{S}M) by the grant from the Research Council of Lithuania (LAT-08/2016).

\end{acknowledgements}

\appendix

\section{Abundance ratios of iron-peak elements}
\label{sec:appendix1}

In addition to our [X/Fe] and [X/Mg] plots (Figs.~\ref{fig:EL-trends} and \ref{fig:EL-trends-mg}), we also provide other interesting combinations of iron-peak abundance ratios as additional
material in Figures~\ref{fig:EL-trends-mn}, \ref{fig:EL-trends-ni}, \ref{fig:EL-trends-cu}, and \ref{fig:EL-trends-zn}.

  \begin{figure}[htb]
   \advance\leftskip 0cm
   \includegraphics[scale=0.45]{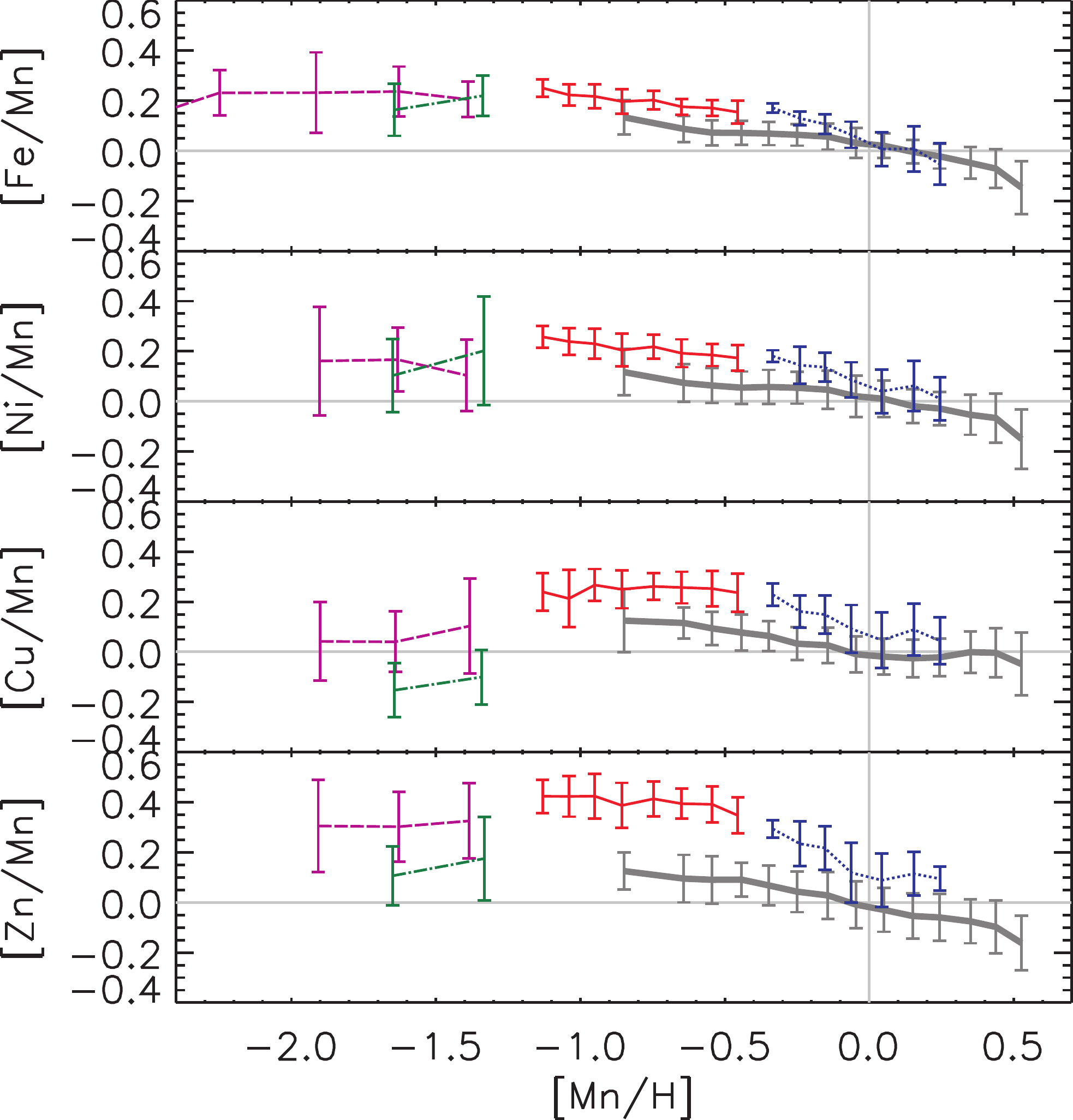}
  \caption{Same plot as Fig.~\ref{fig:EL-trends-mg} but for the [X/Mn] vs. [Mn/H] ratios. The binning structure is the same as in Figs~\ref{fig:EL-scatter}~and~\ref{fig:EL-trends}. The error bars represent the standard deviation associated with the mean value.
}
  \label{fig:EL-trends-mn}
  \end{figure} 

  \begin{figure}[htb]
   \advance\leftskip 0cm
   \includegraphics[scale=0.45]{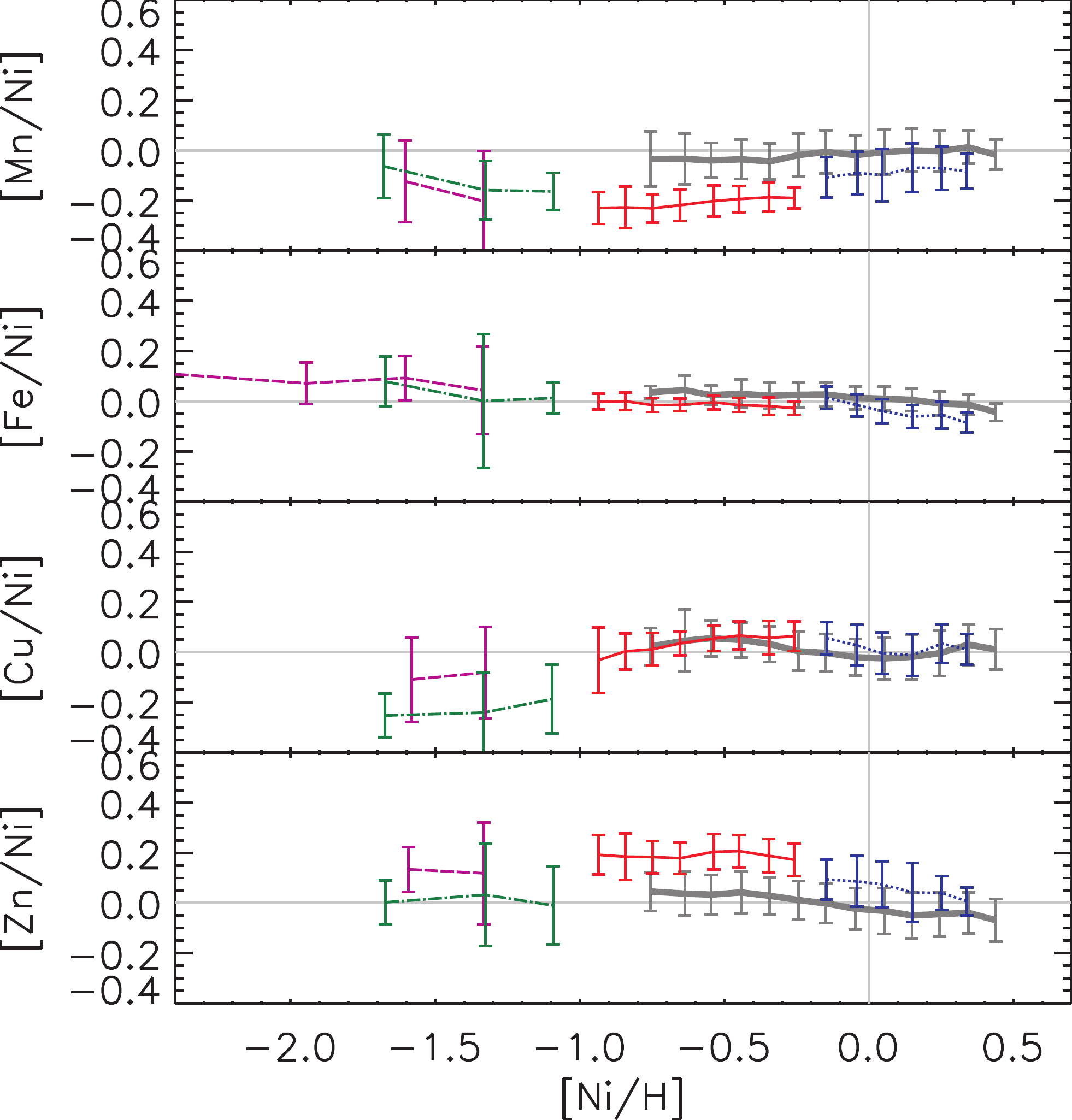}
  \caption{Same plot as Fig.~\ref{fig:EL-trends-mn} but for the [X/Ni] vs. [Ni/H] ratios.}
  \label{fig:EL-trends-ni}
  \end{figure} 
  
  \begin{figure}[htb]
   \advance\leftskip 0cm
   \includegraphics[scale=0.45]{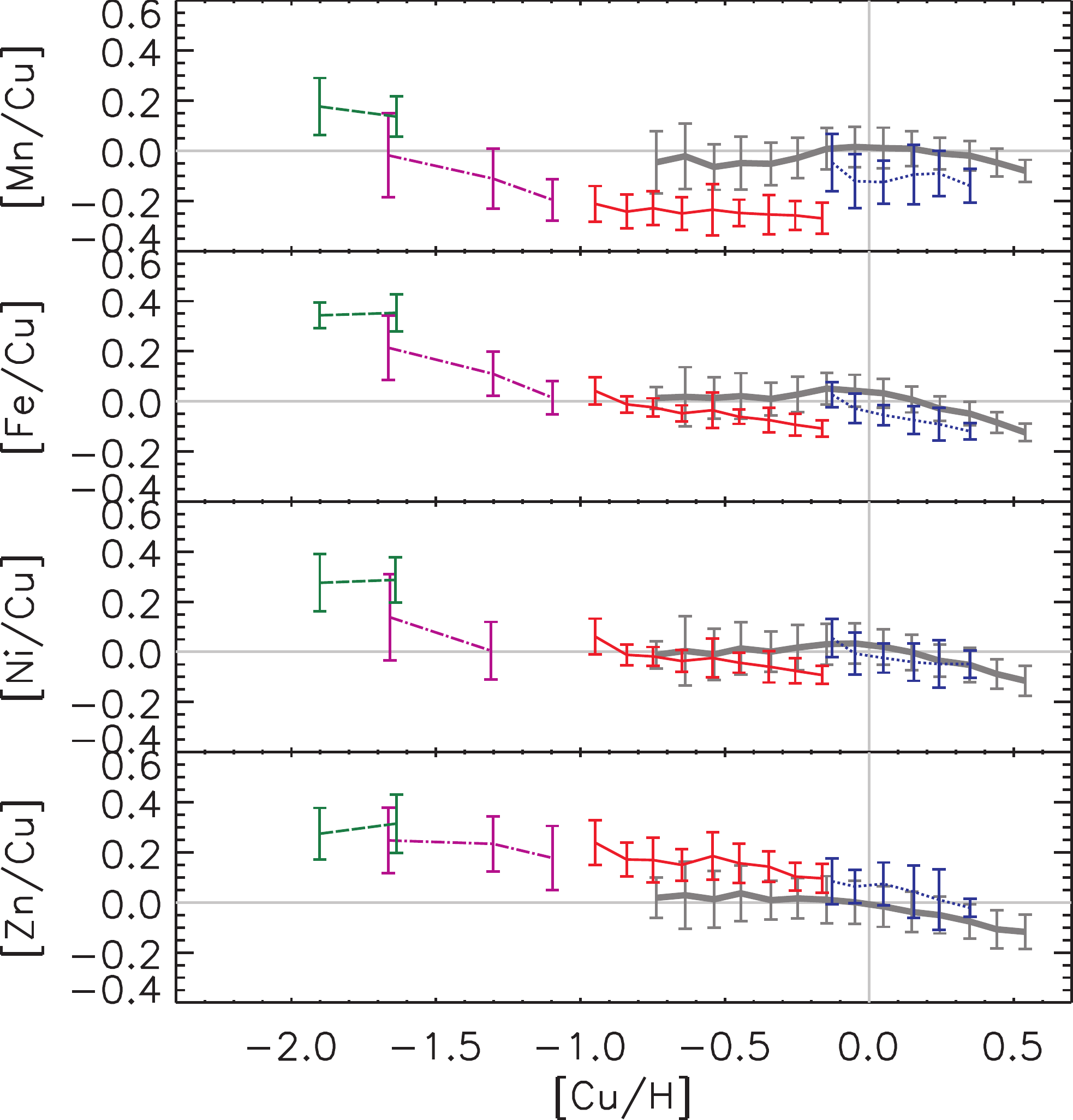}
  \caption{Same plot as Fig.~\ref{fig:EL-trends-mn} but for the [X/Cu] vs. [Cu/H] ratios.}
  \label{fig:EL-trends-cu}
  \end{figure} 
  
  \begin{figure}[htb]
   \advance\leftskip 0cm
   \includegraphics[scale=0.45]{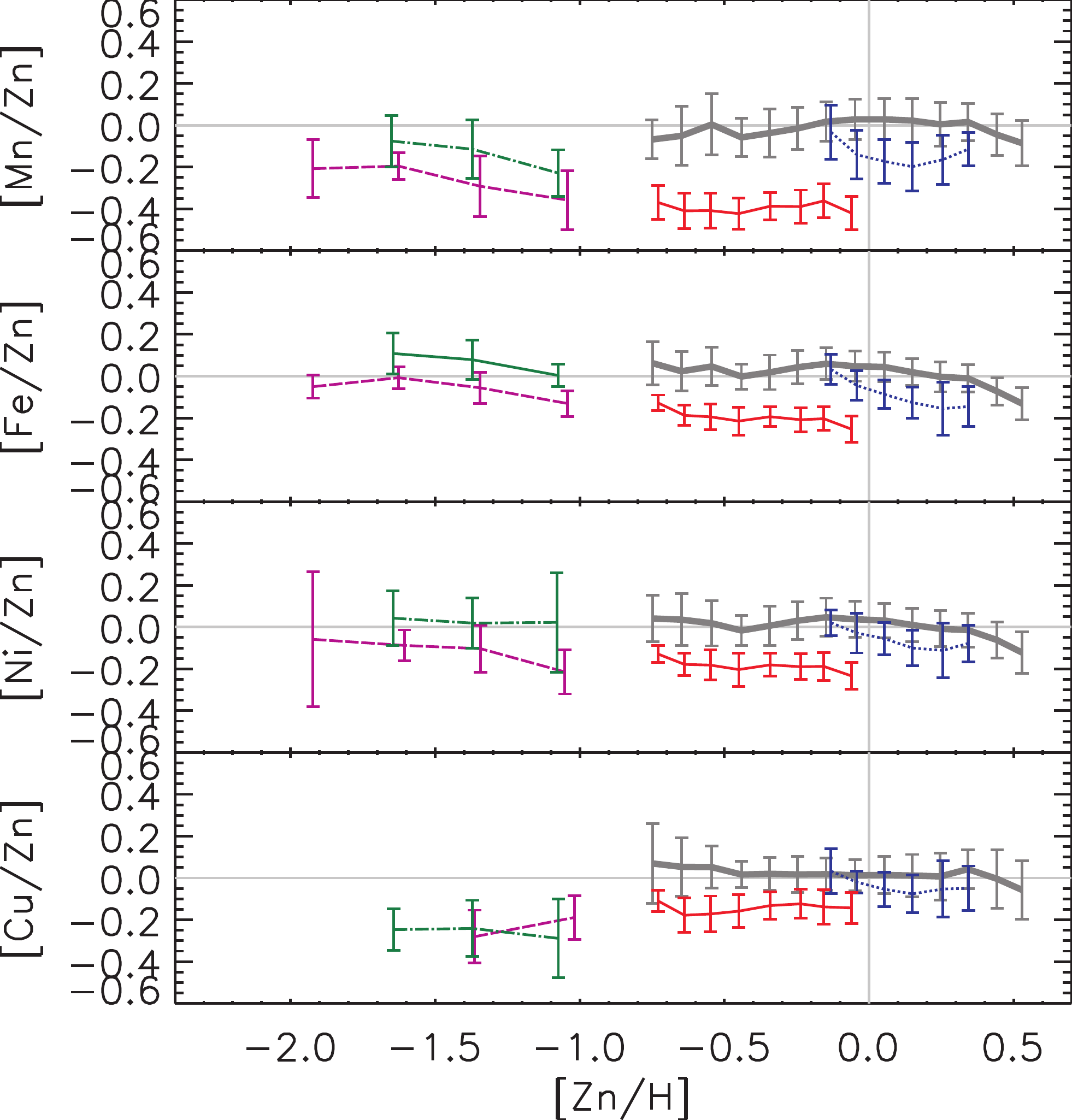}
  \caption{Same plot as Fig.~\ref{fig:EL-trends-mn} but for the [X/Zn] vs. [Zn/H] ratios.}
  \label{fig:EL-trends-zn}
  \end{figure}

\end{document}

%% file: tab-montecarlo.tex
\begin{table}[h]
\caption{Mean and dispersion of atmospheric parameters, signal-to-noise ratios and the derived chemical abundances for repeated spectra. The numbers in parenthesis refers to the number of analysed spectra.}
\begin{tabular}{lcccccc}
\hline\hline
Line & Mean & $\sigma$ & Mean & $\sigma$ \\
\hline
	\multicolumn{5}{c}{HARPS}&\\
            &\multicolumn{2}{c}{HD~188512 (1033$^*$)}&\multicolumn{2}{c}{HD~190248 (1160)}\\
$T_{\rm eff}$          &	5132.56	&	54.26	&	5567.14	&	3.77	\\
${\rm log}~g$            &	   3.53	&	 0.11	&	   4.21	&	0.00	\\
${\rm [M/H]}      $      &	  -0.27	&	 0.01	&	   0.30	&	0.00	\\
${\rm S/N}      $        &	 123.80	&	36.73	&	   99.9	&      38.39	\\
${\rm	[\ion{Mg}{I}/M]}$&	   0.07	&	 0.00	&	   -0.0	&	0.00	\\
${\rm	[\ion{Fe}{I}/M]}$&	  -0.13	&	 0.03	&	  -0.15	&	0.01	\\
${\rm	[\ion{Fe}{II}/M]}$&	   0.04	&	 0.01	&	  -0.01	&	0.00	\\
${\rm	[\ion{Mn}{I}/M]}$&	  -0.02	&	 0.01	&	  -0.08	&	0.00	\\
${\rm	[\ion{Ni}{I}/M]}$&	   0.03	&	 0.01	&	   0.04	&	0.01	\\
${\rm	[\ion{Cu}{I}/M]}$&	   0.01	&	 0.03	&	   0.02	&	0.04	\\
${\rm	[\ion{Zn}{I}/M]}$&	  -0.14	&	 0.04	&	  -0.22	&	0.02	\\
\hline
            &\multicolumn{2}{c}{HD~31128 (47)}&\multicolumn{2}{c}{HD~11977 (181)}\\
$T_{\rm eff}$          &	5807.10	&	37.14	&	4970.80	&      33.10	\\
${\rm log}~g$            &	   3.97	&	 0.05	&	   2.46	&	0.03	\\
${\rm [M/H]}      $      &	  -1.74	&	 0.01	&	  -0.31	&	0.03	\\
${\rm S/N}      $        &	   81.7	&	28.38	&	   52.6	&      54.85	\\
${\rm	[\ion{Mg}{I}/M]}$&	   0.37	&	 0.00	&	   0.17&	0.00	\\
${\rm	[\ion{Fe}{I}/M]}$&	   0.23	&	 0.03	&	   0.16	&	0.01	\\
${\rm	[\ion{Fe}{II}/M]}$&	   0.24	&	 0.01	&	   0.11	&	0.00	\\
${\rm	[\ion{Mn}{I}/M]}$&	  -0.32	&	 0.01	&	   0.19	&	0.00	\\
${\rm	[\ion{Ni}{I}/M]}$&	  -0.09	&	 0.01	&	   0.12	&	0.01	\\
${\rm	[\ion{Cu}{I}/M]}$&	  -0.31	&	 0.03	&	   0.28	&	0.04	\\
${\rm	[\ion{Zn}{I}/M]}$&	   0.17	&	 0.04	&	   0.15	&	0.02	\\
\hline
	\multicolumn{5}{c}{FEROS}&\\
            &\multicolumn{2}{c}{HD~217107 (107)}&\multicolumn{2}{c}{HD~156384B (38)}\\

$T_{\rm eff}$          &	5624.31	&	 7.49	&	4648.78	&	 8.90	\\
${\rm log}~g$            &	   4.44	&	 0.00	&	   4.60	&	 0.01	\\
${\rm [M/H]}      $      &	   0.25	&	 0.01	&	  -0.47	&	 0.00	\\
${\rm S/N}      $        &	  66.26	&	22.97	&	  96.75	&      	20.01	\\
${\rm	[\ion{Mg}{I}/M]}$&	   0.18	&	 0.02	&	   0.16	&	 0.01	\\
${\rm	[\ion{Fe}{I}/M]}$&	   0.17	&	 0.02	&	   0.04	&	 0.01	\\
${\rm	[\ion{Fe}{II}/M]}$&	   0.17	&	 0.01	&	   0.04	&	 0.01	\\
${\rm	[\ion{Mn}{I}/M]}$&	   0.12	&	 0.02	&	   0.04	&	 0.04	\\
${\rm	[\ion{Ni}{I}/M]}$&	   0.23	&	 0.01	&	   0.06	&	 0.01	\\
${\rm	[\ion{Cu}{I}/M]}$&	   0.23	&	 0.01	&	   0.12	&	 0.01	\\
${\rm	[\ion{Zn}{I}/M]}$&	   0.08	&	 0.02	&	   0.11	&	 0.04	\\
\hline
           &\multicolumn{2}{c}{HIP~11952 (26)}&\multicolumn{2}{c}{HD~10042 (25)}\\
$T_{\rm eff}$          &	5866.41	&	15.68	&	4903.90	&	20.04	\\
${\rm log}~g$            &	   3.65	&	 0.05	&	   2.59	&	 0.06	\\
${\rm [M/H]}      $      &	  -1.93	&	 0.04	&	  -0.44	&	 0.05	\\
${\rm S/N}      $        &	  78.26	&	28.75	&	  67.34	&      	30.50	\\
${\rm	[\ion{Mg}{I}/M]}$&	   0.43	&	 0.02	&	   0.51	&	 0.01	\\
${\rm	[\ion{Fe}{I}/M]}$&	   0.35	&	 0.02	&	   0.21	&	 0.01	\\
${\rm	[\ion{Fe}{II}/M]}$&	   0.27	&	 0.01	&	   0.12	&	 0.01	\\
${\rm	[\ion{Mn}{I}/M]}$&	   0.12	&	 0.02	&	   0.03	&	 0.04	\\
${\rm	[\ion{Ni}{I}/M]}$&	  -0.05	&	 0.01	&	   0.16	&	 0.01	\\
${\rm	[\ion{Cu}{I}/M]}$&	  -0.23	&	 0.01	&	   0.26	&	 0.01	\\
${\rm	[\ion{Zn}{I}/M]}$&	   0.32	&	 0.02	&	   0.41	&	 0.04	\\
\hline
\hline
\label{tab:Montecarlo}
\end{tabular}

$^*$ Number of repeats.
\end{table}

%% file: tab-sensitivity.tex
   \begin{table}
\begin{center}

      \caption{Effects on the derived abundances ([X/H]), resulting from the atmospheric parameters uncertainties$^4$
for four types of stars, selected by $T_{\rm eff}$, and ${\rm log}~g$.
}
          \label{tab:Sensitivity}
      \[
      \resizebox{\columnwidth}{!}{%
         \begin{tabular}{lrrccccc}
            \hline
            \noalign{\smallskip}
	    El & 
	    ${ \Delta T_{\rm eff} }$ & 
            ${ \Delta \log g }$ & 
            $\Delta {\rm [M/H]}$ & 
            ${ \Delta v_{\rm t} }$ 
	     & $ \sigma_{\rm total\left[\frac{X}{H}\right]} $  
     	     & $ \sigma_{\rm total\left[\frac{X}{Fe}\right]} $  
	     & $ \sigma_{\rm all\left[\frac{X}{Fe}\right]} $  
	     \\ 
            \noalign{\smallskip}
            \hline
            \noalign{\smallskip}
\multicolumn{5}{c}{$T_{\rm eff}<5000    K$, ${\rm log}~g<3.5   $}\\
Mg\,{\sc i} 	&	0.08	&	0.11	&	0.03	&	0.03	&	0.15	&   0.08 & 0.10	\\
Mn\,{\sc i} 	&	0.06	&	0.01	&	0.01	&	0.03	&	0.07	&   0.04 & 0.05	\\
Fe\,{\sc i} 	&	0.05	&	0.06	&	0.07	&	0.02	&	0.11	&   0.06 & 0.07	\\
Fe\,{\sc ii} 	&	0.08	&	0.11	&	0.02	&	0.04	&	0.15	&   0.08 & 0.10	\\
Ni\,{\sc i} 	&	0.03	&	0.04	&	0.02	&	0.02	&	0.06	&   0.03 & 0.04	\\
Cu\,{\sc i} 	&	0.12	&	0.10	&	0.08	&	0.04	&	0.18	&   0.10 & 0.12	\\
Zn\,{\sc i} 	&	0.11	&	0.09	&	0.09	&	0.04	&	0.17	&   0.10 & 0.11	\\
                 \noalign{\smallskip}
\hline
                 \noalign{\smallskip}
\multicolumn{5}{c}{$T_{\rm eff}>5000    K$, ${\rm log}~g<3.5   $}\\
Mg\,{\sc i} 	&	0.07	&	0.10	&	0.02	&	0.03	&	0.12	&   0.07  &   0.08	\\
Mn\,{\sc i} 	&	0.07	&	0.01	&	0.01	&	0.03	&	0.08	&   0.04  &   0.05	\\
Fe\,{\sc i} 	&	0.07	&	0.10	&	0.09	&	0.02	&	0.16	&   0.09  &   0.10	\\
Fe\,{\sc ii} 	&	0.02	&	0.08	&	0.01	&	0.04	&	0.09	&   0.05  &   0.06	\\
Ni\,{\sc i} 	&	0.07	&	0.02	&	0.01	&	0.02	&	0.08	&   0.04  &   0.05	\\
Cu\,{\sc i} 	&	0.08	&	0.02	&	0.02	&	0.04	&	0.09	&   0.05  &   0.06	\\
Zn\,{\sc i} 	&	0.03	&	0.04	&	0.03	&	0.04	&	0.08	&   0.04  &   0.05	\\
                 \noalign{\smallskip}
\hline
                 \noalign{\smallskip}\multicolumn{5}{c}{$T_{\rm eff}<5000    K$, ${\rm log}~g>3.5   $}\\
Mg\,{\sc i} 	&	0.05	&	0.09	&	0.04	&	0.04	&	0.11	&   0.06  &   0.07	\\
Mn\,{\sc i} 	&	0.03	&	0.03	&	0.01	&	0.03	&	0.05	&   0.03  &   0.03	\\
Fe\,{\sc i} 	&	0.07	&	0.05	&	0.08	&	0.02	&	0.12	&   0.07  &   0.08	\\
Fe\,{\sc ii} 	&	0.10	&	0.09	&	0.02	&	0.03	&	0.14	&   0.08  &   0.09	\\
Ni\,{\sc i} 	&	0.01	&	0.05	&	0.02	&	0.03	&	0.06	&   0.03  &   0.04	\\
Cu\,{\sc i} 	&	0.03	&	0.06	&	0.02	&	0.04	&	0.08	&   0.04  &   0.05	\\
Zn\,{\sc i} 	&	0.06	&	0.05	&	0.04	&	0.04	&	0.10	&   0.06  &   0.07	\\
                 \noalign{\smallskip}
\hline
                 \noalign{\smallskip}\multicolumn{5}{c}{$T_{\rm eff}>5000    K$, ${\rm log}~g>3.5   $}\\
Mg\,{\sc i} 	&	 0.04	&	0.04	&	0.03	&	0.03	&	0.09	&  0.05 &  0.06	\\
Mn\,{\sc i} 	&	 0.03	&	0.03	&	0.01	&	0.02	&	0.04	&  0.02 &  0.03	\\
Fe\,{\sc i} 	&	 0.06	&	0.05	&	0.07	&	0.02	&	0.09	&  0.05 &  0.06	\\
Fe\,{\sc ii} 	&	 0.09	&	0.07	&	0.02	&	0.03	&	0.11	&  0.06 &  0.07	\\
Ni\,{\sc i} 	&	 0.01	&	0.03	&	0.02	&	0.03	&	0.06	&  0.03 &  0.04	\\
Cu\,{\sc i} 	&	 0.03	&	0.05	&	0.02	&	0.03	&	0.07	&  0.04 &  0.05	\\
Zn\,{\sc i} 	&	 0.05	&	0.05	&	0.03	&	0.04	&	0.09	&  0.05 &  0.06	\\
            \hline
         \end{tabular}}
      \]
  \end{center}
 
The table shows the median of the propagated errors on [X/H] ratios due to $\Delta T_{\rm eff}$, $\Delta {\rm log}~g$, $\Delta {\rm [M/H]}$, $\Delta v_{\rm t}$.
The $ \sigma_{\rm total([X/H])} $ stands for the median of the quadratic sum of all four effects on [X/H] ratios. 
The $ \sigma_{\rm total([X/Fe])} $ stands for the median of the quadratic sum  of all four effects on [X/Fe] ratios. 
The $ \sigma_{\rm all([X/Fe])} $ is the combined effect of $ \sigma_{\rm total([X/Fe])} $ and the line-to-line scatter from Table~\ref{tab:Scatter}. 


%
   \end{table}